\documentclass[11pt,a4paper]{article}
\usepackage[utf8]{inputenc}
\usepackage[english]{babel}
\usepackage{amsmath}
\usepackage{amsfonts}
\usepackage{amssymb}
\usepackage{graphicx}
\usepackage{lmodern}
\usepackage{booktabs}
\usepackage{authblk}
\usepackage{url}
\usepackage{hyperref}
\usepackage{verbatim}
\usepackage{color}
\usepackage{comment}
\usepackage{float}
\usepackage[usenames,dvipsnames]{xcolor}
\hypersetup{colorlinks=true, citecolor=black, urlcolor=black, linkcolor=black}

\makeatletter
\renewcommand*\@fnsymbol[1]{\the#1}
\makeatother
\usepackage[left=2cm,right=2cm,top=2.5cm,bottom=2cm]{geometry}

\title{Collaboration and followership: \\
a stochastic model for activities in social networks}

\author{ Carolina Becatti, Irene Crimaldi, Fabio
  Saracco\footnote{Alphabetic order.  E-mails:
    carolina.becatti@imtlucca.it (corresponding author),
    irene.crimaldi@imtlucca.it, fabio.saracco@imtlucca.it}\\ {\em
    \small IMT School for Advanced Studies Lucca, Piazza San Ponziano
    6, 55100 Lucca, Italy} }

\date{}

\begin{document}

\maketitle
\setcounter{footnote}{1}

\begin{abstract}
\noindent 
In this work we investigate how future actions are 
influenced by the previous ones, in the specific contexts of scientific 
collaborations and friendships on social networks.
We are not interested in modeling the process of link
formation between the agents themselves, we instead describe the
activity of the agents, providing a model for the formation of the
{\em bipartite network of actions and their features}. Therefore
we only require to know the chronological order in which the actions
are performed, and not the order in which the agents are
observed. Moreover, the total number of possible features is not
specified a priori but is allowed to increase along time, and new
actions can independently show some new-entry features or exhibit some
of the old ones.  The choice of the old features is driven by a
degree-fitness method.  With this term we mean that the probability
that a new action shows one of the old features does not solely depend
on the ``popularity'' of that feature (i.e.~the number of previous
actions showing it), but is also affected by some individual traits of
the agents or the features themselves, synthesized in certain
quantities, called ``fitnesses'' or ``weights'', that can have
different forms and different meaning according to the specific
setting considered. We show some theoretical properties of the model
and provide statistical tools for the parameters' estimation.  The
model has been tested on three different datasets and the numerical
results are provided and discussed.
\\

\noindent{\em keywords:} bipartite networks, preferential attachment,
fitness, collaboration networks, followership networks, on-line social
networks, arXiv, IEEE, Instagram.

\end{abstract}

\section{Introduction}
\label{intro}
In the last years complex networks established as a proper tool for
the description of the interactions within large
systems~\cite{caldarelli2010scale-free,NEWMAN2010}. The renewed
attention to this field can be dated back to the well known
Barab\'{a}si-Albert model~\cite{Barabasi1999}, in which the authors
provide an explanation of the power-law distribution of node degrees
in the World Wide Web (WWW) via a dynamic generative network model. At
every step a new vertex is added and the probability to observe a new
link is proportional to the number of connections (i.e. the degree) of
the target node. The success of this proposal resides in the fact that
only this simple rule, called \emph{preferential attachment}, is able
to reproduce with good accuracy the degree distribution of many real
networks, such as the WWW.  Even if the original mechanism was already
present in the literature in a slightly different
form~\cite{DeSollaPrice1965,Yule1925}, the paper of
Bar\'{a}basi-Albert boosted the attractiveness of complex networks and
other scholars delved into the investigation of the properties of
generative models (nice reviews on the subject
are~\cite{Golosovsky2018,Newman2005}). In the
articles~\cite{Krapivsky2001,Krapivsky2000}, the effects of having
connection probabilities proportional to a positive power of the
degrees is considered: probabilities per link less than linear produce
an exponential degree distribution, while those more than linear
produce the emergence of a completely connected node. The preferential
attachment was then enriched with another ingredient, such as the
\emph{fitness}~\cite{Bianconi2001,Bianconi2001a}: a quantity defined
per node that measures the intrinsic ability of the vertex to collect
links. Then, the probability of targeting a certain node becomes the
product of its fitness and degree.  The effect of this new variable is
to amplify or dampen the preferential attachment effect.  Indeed, the
presence of the fitness permits to overcome the ``first move
advantage'' (i.e. the fact that older nodes have greater degrees by
construction), thus permitting to ``young" nodes to grow
easily. Beside generative models, the node fitness can be generalised
to describe the structure of real networks by correlating its value to
some attributes of the nodes, not directly specified in the definition
of the network~\cite{Caldarelli2002}. In a recent
paper~\cite{Golosovsky2018}, the proposal of~\cite{Caldarelli2002} was
extended to build a generative model that solely embeds fitnesses and
not node degrees: by modifying their distribution, fitnesses only are
able to reproduce the power-law degree distributions present in many
networks. Thus, which should be the fundamental quantity for the
description of the network, either node's degree or fitness, is
argument of debate~\cite{Golosovsky2018}.\\ \indent Time dependence is
generally included by considering the possibility of node ageing,
i.e. multiplying the probability of link by a time dependent damping
function~\cite{Dorogovtsev2000,Golosovsky2018,
  Medo2011,Wang2013}. In~\cite{Dorogovtsev2000} the original
preferential attachment is modified by introducing an ageing factor
proportional to a power-law of the age of the target node. By
modifying the exponent of the ageing factor, the authors recover
different power-law distributions for the degree sequence: if the
ageing exponent is negative, the exponent for the degree distribution
is smaller than its analogous for non-ageing preferential
attachment. Instead, for positive ageing exponent, the degree
distribution's exponent increases, eventually turning the distribution
to an exponential one. In~\cite{Medo2011} node ageing is captured by a
fitness that decays with time: the resulting degree distributions may
be exponential, log-normal or power-law, depending of the fitness
definition. Finally, in order to quantify the impact of scientific
production,~\cite{Wang2013} proposed a probability per link that
comprehends a (static) fitness, the degree and a time dependence
factor.\\ \indent The importance of the previous proposals was not in
the definition of the model per se, but in providing an explanation
for the structure of the networks examined. For instance, the
preferential attachment in~\cite{Barabasi1999} explains the power-law
degree distribution in the World Wide Web and describes a ``rich get
richer'' competition for links. Instead, in the fitness methods, some
attributes of the nodes, not directly observed in the network, define
the structure of the network (as in the case of e-mails networks, in
which senders do not have access to information about the number of
connection of the receivers~\cite{Caldarelli2002}). In the same way,
fitness ageing~\cite{Medo2011} gives an explanation to the limited (in
time) growth in citation of most of the papers.\\ \indent All previous
efforts were devoted to monopartite, directed or undirected,
networks. A much smaller number of contributions is available for the
description of the evolution of {\em bipartite networks}. In bipartite
networks, nodes are divided into two different classes, called
``layers'', and only links connecting nodes belonging to different
layers are allowed~\cite{caldarelli2010scale-free,NEWMAN2010}.
Guillame and Latapy~\cite{Guillaume2006} proposed a simple model:
consider the case in which the degree distribution on one layer is
given and is power-law on the other one.  Then sample a certain degree
$d$ for a node on the former layer and connect it to $d$ existing
nodes on the opposite layer, selected with a preferential attachment
procedure. In case both layers have a power-law degree distribution
(like reviews and reviewers in the Netflix dataset), these
distributions can be reproduced adding one single link at every time
step, selecting nodes on each layer by mixing uniform and preferential
attachment~\cite{BeguerisseDiaz2010}.  Some other dynamical models for
bipartite networks were proposed for the description of specific
systems. For instance, in~\cite{calvao2015consumer} the authors
propose a generative model to study the bipartite networks of lawyers
and clients that develops according to a recommendation process: more
popular lawyers are also more likely to be hired by new
clients. Furthermore, the authors in~\cite{roth2010social} provide a
framework in which the simultaneous evolution of two systems has been
studied.  Indeed, they analyse communities of scientists considering
both the monopartite network describing the interactions among agents
themselves and the bipartite semantic network in which the agents are
associated to the concepts they use. Another example
is~\cite{Saracco2015}, in which the structure of the (growing)
bipartite trade network (layers represent countries and exported
products) was reproduced by assigning links with sequential
preferential attachment, considering the degree of both nodes in the
process.  In order to describe the generation of an innovative
product, following the idea of the ``adjacent
possibles''~\cite{Kauffman1995}, new nodes (i.e. new products) are
derived by the structure of an unobserved mono-partite network of
products describing the hierarchical productive process relations.
Therefore, the evolution of the bipartite system is due to the
simultaneous dynamics of an unobserved evolving network. \\ \indent In
order to define a network model based on a latent attribute structure,
a new model was introduced in \cite{Boldi2016}. In this context, a set
of nodes sequentially join the considered network, each of them
showing a set of features. Each node can either exhibit new features
or adopt some of the features already present in the network. This
choice is regulated by a preferential attachment rule: the larger the
number of nodes showing a certain feature, the greater the probability
that future nodes will adopt it too. The total number of possible
features is not specified a priori, but is allowed to increase along
time. Differently from~\cite{Birmele2009,Guillaume2006}, each node has
been weighted with a fitness variable, that accounts for nodes'
personal ability to transmit its own features to future
nodes. Starting from here, the model in \cite{CRIMALDI2017} introduces
some novelties in the previous context: the probability to exhibit one
of the features already present in the network is defined as a
mixture, i.e.~a convex combination, of random choice and preferential
attachment. However, neither fitnesses nor weights are introduced in
the model, so that all nodes are assumed to have equal capabilities in
transmitting their personal features to the newcomers.  \\ \indent The
present work moves along the same research line of the previously
mentioned papers~\cite{Boldi2016,CRIMALDI2017}, but with a different
spirit.  First of all, the previous papers provide two different
models of network formation, in which the nodes sequentially join the
network and the number of common features affects the probability of
connections among them.  The main drawback of these two models resides
in the assumed chronological order of nodes' arrivals, which may
tipically be unknown (or non-relevant) in many real-world systems.  In
the present paper we overcome this limitation: given a system of $n$
agents, we provide a {\em model for the formation of the bipartite
  network of agents' actions and their features}. This model can also
be applied to all settings in which agents of interest are not
observed in a specific chronological order, because the assumption on
the chronological order is specified on the agents' actions
only. Furthermore the probability to exhibit one of the features
already observed is defined as a mixture of random choice and {\em
  ``preferential attachment with weights''}, i.e.  the probability of
connection depends both on the features' degrees and the fitness of
the agents involved and/or of the features themselves.  These weights
$W_{t,j,k}$ can have different forms and meanings according to the
specific setting considered: the weight at time-step $t$ of the
observed feature $k$ can depend on some characteristics of $k$ itself,
or it can be directly established by the agent performing 
action $t$;
it may also represent the ``inclination'' of the agent performing action
$t$ in adopting the previous observed features, or 
some properties of the agent performing the previous action $j$
with $k$ among its features (for instance, her/his ability to transmit
her/his own features).  \\ \indent We analyse two datasets of
scientific publications (respectively IEEE for Automatic Driving, and
arXiv for Theoretical High Energy Physics, or more briefly Hep-Th) and
a dataset of posts of Instagram.  We not only obtain a good fit
of our model to the data, but our analysis also results useful in
order to highlight interesting aspects of the activity of the three
considered social networks. Indeed, we find different variables 
playing a role in their evolution.  In the three
  systems studied, we consider the degrees of the features (i.e. the
  popularity of, respectively, keywords in a scientific paper or
  hashtags on Instagram) and some fitness variables associated to the
  agents as drivers for the dynamics. For the scientific publications,
  we show a good agreement of the model to the IEEE dataset for
  Automatic Driving and to the arXiv dataset for Hep-Th with weights
  based on the number of publications or the number of co-authors of
  an author, the former performing better in the case of Automatic
  Driving. Otherwise stated, in the case of Automatic Driving the
  ability of an author to transmit the keywords of her/his papers,
  that essentially describe her/his research topics, is better
  reproduced by her/his number of publications, while in Hep-Th this
  ability is related both to the activity of the author, i.e. to the
  number of her/his publications, and to the number of collaborations
  established in her/his career. This difference can be due to the
  nature of the two research fields. Automatic Driving is more recent
  and limited, and new results are driving the evolution of the
  research. Thus, an author transmits more keywords the more its
  activity in the research.  Hep-Th research area, instead, is an
  older and structured research field, evolved in different
  specialised branches.
In the case of on-line social networks, the evolution is, instead,
guided by the popularity of the users, but in a tricky sense: a
standard user tends to follow many already existing hashtags, in order
to acquire more visibility, while famous users mention just few
hashtags, already being popular.\\ \indent The present paper is so
organized. In Section \ref{model} we illustrate in detail the proposed
model for the formation of the actions-features bipartite network.  In
Section \ref{meaning} we explain the meaning of the model parameters
and the role of the weights introduced into the preferential
attachment term.  Some asymptotic results regarding the behavior of
the total number of features and the mean number of edges in the
actions-features bipartite network are collected in the Appendix,
Subsection \ref{appendix-asymptotics}. The Appendix also contains a
description of the statistical tools for the estimation of the model
parameters (see Subsection \ref{estimation}).  In Section
\ref{applications} we briefly provide the general methodology used to
analyse the data (the details are postponed in the Appendix,
Subsection \ref{methodology}), and then we show the application of our
model to the above mentioned real-world cases (IEEE, arXiv, Instagram
datasets).  We summarize the overall contents of the paper and recap
the main obtained findings in the last Section \ref{conclusion}.

\section{Model for the dynamics of the actions-features network}
\label{model}

Suppose to have a system of $n$ agents that sequentially perform
actions along time. Each agent can perform more than one action. The
running of the time-steps coincide with the flow of the actions and so
sometimes we use the expression ``time-step $t$'' in order to indicate
the time of action $t$. Each action is characterized by a finite
number of features and different actions can share one or more
features. It is important to point out that we do not specify a priori
the total number of possible features in the system, but we allow this number 
to increase along time. In what follows, we describe
the model for the dynamical evolution of the bipartite network that
collects actors' actions on one side and the corresponding features of
interest on the other side. We denote by $F$ the adjacency matrix
related to this network. The dynamics starts with the observation of
action $1$, the first action done by an agent of the considered
system, that shows $N_1$ features, where $N_1$ is assumed Poisson
distributed with parameter $\alpha>0$. (This distribution will be
denoted from now on by the symbol Poi$(\alpha)$). Moreover, we number
the observed features with $k$ from $1$ to $N_1$ and we set
$F_{1,k}=1$ for $k=1,\dots, N_1$. Then, for each consecutive action $t
\geq 2$, we have:
\begin{enumerate}
\item Action $t$ exhibits some old features, where ``old'' means
  already shown by some of the previous actions $1, \dots, t-1$.  More
  precisely, if $N_j$ denotes the number of new features exhibited by
  action $j$ and we set
\begin{equation}
L_{t-1}=\sum_{j=1}^{t-1}N_j=\hbox{the overall number of different
  observed features for the first $t-1$ actions},
\end{equation}
 the new action $t$ can independently display each old feature $k \in
 \{1, \dots, L_{t-1}\}$ with probability
\begin{equation}\label{incl_prob}
P_t(k) = \frac{\delta}{2} + 
(1-\delta)\frac{\sum_{j=1}^{t-1}F_{j,k}W_{t,j,k}}{B_t}
\end{equation}
where $\delta\in [0,1]$ is a parameter, $F_{j,k} = 1$ if action $j$
shows feature $k$ and $F_{j,k} = 0$ otherwise, $W_{t,j,k}\geq 0$ is
the random weight associated to feature $k$ measured at the time of 
action $t$ that can be related to the course of previous actions $j$.
Finally $B_t$ is a suitable normalizing factor so that
$\sum_{j=1}^{t-1}F_{j,k}W_{t,j,k}/B_t$ belongs to $[0,1]$. We will
refer to quantity \eqref{incl_prob} as the ``inclusion probability'' of
feature $k$ at time-step $t$.
\item Action $t$ can also exhibit a number of new features $N_t$,
  where $N_t$ is assumed Poi$(\lambda_t)$-distributed with parameter
\begin{equation}\label{lambda}
\lambda_t = \frac{\alpha}{t^{1-\beta}}, 
\end{equation}
where $\beta\in [0,1]$ is a parameter. The variable $N_t$ is supposed
independent of $N_1, \dots, N_{t-1}$ and of all the appeared old
features and their weights (including those of action $t$).  
\end{enumerate}
With the observation of the $t^{th}$ action, all the matrix elements 
$F_{t,k}$ with $k \in \{1, \dots, L_t\}$ are set equal to $1$ if action 
$t$ shows feature $k$ and equal to $0$ otherwise. Here is an example 
of a $F$ matrix with $t=3$ actions:
\[
F=\left(
\begin{array}{cccccccccccc}
{\bf 1} & {\bf 1 } & {\bf 1} & {\bf 1} & 0 & 0 & 0 & 0 & 0
\\
1 & 0 & 1 & 0 & {\bf  1} & {\bf 1} & 0 & 0 & 0
\\
1 & 0 & 1 & 1 & 0 & 1 & {\bf 1} & {\bf 1} & {\bf 1}\\
\end{array}
\right).
\]
\noindent In boldface we highlight the new features for each action:
we have $N_1=4$, $N_2=2$, $N_3=3$ and so $L_1=4,L_2=6,L_3=9$ and, for
each action $t$, we have $F_{t,k}={\bf 1}$ for each $k \in
\{L_{t-1}+1,\dots,L_t\}$. Moreover, some elements $F_{t,k}$, with $k
\in \{1,\dots,L_{t-1}\}$, are equal to $1$ and they represent the
features brought by previous actions exhibited also by action $t$.
\\

\indent It may be worth to note that our model resembles the one known
as the ``Indian buffet process'' in Bayesian Statistics
\cite{berti2015central,ghahramani2006infinite,teh2009indian}, but
indeed there are significant differences in the definition of the
inclusion probabilities: in particular, the mixture parameter $\delta$
and the weights $W_{t,j,k}$. Moreover, Bayesian Statistics deals with
exchangeable sequences, while here we do not require this property. As
a consequence, the role played by each parameter in \eqref{incl_prob} 
and \eqref{lambda} results more straightforward and easy to be
implemented.

\section{Discussion of the model}
\label{meaning}

We now discuss the meaning of the model parameters $\alpha,\, \beta$
and $\delta$ and the role of the random weights $W_{t,j,k}$. Some
asymptotic results are collected in the Appendix, Subsection
\ref{appendix-asymptotics}; while the statistical tools employed to
estimate the model parameters are provided in the Appendix,
Subsection \ref{estimation}.

\subsection{The parameters $\alpha$ and $\beta$}
\label{alpha-beta}
In the above model dynamics, the probability distribution of the
random number $N_t$ of new features brought by action $t$ is regulated
by the pair of parameters $(\alpha,\, \beta)$ (see \eqref{lambda}).
Specifically, the larger $\alpha$, the higher the total number of new
features brought by an action, while $\beta$ controls the asymptotic
behavior of the random variable $L_t=\sum_{j=1}^t N_j$, i.e. the total
number of features observed for the first $t$ actions, as a function
of $t$. In particular, it has been shown in \cite{CRIMALDI2017} that
the parameter $\beta>0$ corresponds to the power-law exponent of
$L_t$: precisely, if $\beta = 0$ then the asymptotic behavior of
$L_t$ is logaritmic, while for $\beta \in (0, 1]$ we obtain a
  power-law behavior with exponent $\beta$ (see Subsection
  \ref{total-num-features} in the Appendix).

\subsection{The parameter $\delta$ and the random weights $W_{t,j,k}$} 
\label{delta-weights}

Looking at equation \eqref{incl_prob} of the above model dynamics, we
can see that, for a generic action $t$, both the parameter $\delta$
and the random weights $W_{t,j,k}$ affect the number of old features
($k=1,\dots, L_{t-1}$) also shown by action $t$. Specifically, the
value $\delta = 1$ corresponds to the ``pure i.i.d. case'' with
inclusion probability equal to $1/2$: an action can exhibit each
feature with probability $1/2$ independently of the other actions and
features. The value $\delta = 0$ corresponds to the case in which the
inclusion probability $P_t(k)$ entirely depends on the (normalized)
total weight associated to feature $k$ at the time of action $t$,
i.e. to the quantity
\begin{equation}\label{wpa-factor}
\frac{\sum_{j=1}^{t-1} F_{j,k} W_{t,j,k}}{B_t}.
\end{equation}
In equation (\ref{wpa-factor}), the term $W_{t,j,k}\geq 0$ is
the random weight at time-step $t$ associated to feature $k$ that can be 
related to the course of previous actions $j$. We denote this case as the
``pure weighted preferential attachment case'' since the larger the
total weight of feature $k$, the greater the probability that also the
new action will show feature $k$. When $\delta \in (0, 1)$, we have a
mixture of the two cases above: the smaller $\delta$, the more
significant is the role played by the weighted preferential attachment
in the spreading of the observed features to the new actions. In the
sequel we will refer to \eqref{wpa-factor} as the ``weighted
preferential attachment term''. \\

\indent Regarding the weights, the possible ways in which they can be
defined benefit of a great flexibility. Of course their meaning has to
be discussed in relation to the particular application considered. For
instance, the weight $W_{t,j,k}$ can be ``directly'' assigned by the
agent performing action $t$ to the feature $k$ in connection with the
previous action $j$, or it may represent the ``inclination'' of the
agent performing action $t$ of ``adopting'' the previous observed
features, or it may ``implicitly'' due to some properties of the agent
performing the previous action $j$ (for instance, her/his ability to
transmit her/his own features), or even more. We here describe some
general interesting frameworks:

\begin{itemize} 
\item[1)] If we set $W_{t,j,k}=1$ for all $t,j,k$ with normalizing
  factor $B_t=t$, then all the observed features have the same
  weight. Then the sum in the numerator of \eqref{wpa-factor}
  becomes the ``popularity'' of feature $k$, that is the total number
  of previous actions that have already exhibited feature $k$, while the
  quantity \eqref{wpa-factor} is essentially the ``average
  popularity'' of feature $k$ (we divide by $t$ instead of $t-1$ in
  order to avoid the quantity \eqref{wpa-factor} to be exactly equal
  to $1$ for all the first $N_1$ features). In this case the
  actions-features dynamics coincides with the nodes-features dynamics
  considered in \cite{CRIMALDI2017}.

\item[2)] We can assume that a positive random variable $G_i$ (with
  $i=1,\dots, n$) is associated to each agent in order to describe
  her/his ability to transmit the features of her/his actions to the
  others. This random variable can be seen as a static ``fitness'' as
  defined in~\cite{Bianconi2001, Bianconi2001a, Caldarelli2002}.  In
  this case the weight $W_{t,j,k}$ can be defined as $G_{i(j)}$ (or a
  function of this quantity), where $i(j)$ denotes the agent
  performing action $j$. In particular, we have $W_{t,j,k}=W_j$, that is
  the weights only depend on $j$. Hence, the weight of a feature $k$
  is only due to the fitness of the agent that performs an action with
  $k$ among its features and the sum in the numerator of
  \eqref{wpa-factor} becomes the total weight of the feature $k$ due
  to the agents that have previously exihibited it in their
  actions. The quantity $B_t=c+\sum_{h=1}^{t-1}W_h$ can be chosen as
  normalizing factor, i.e. we basically normalize by the total fitness
  of the agents that have performed actions $1,\dots, t-1$. Note that
  case 1) can be seen as a special case of the present, taking $G_i=1$
  and $c=1$. Moreover, another interesting element to observe is that 
  the weighted
  preferential attachment term \eqref{wpa-factor} can be explained
  with an urn process. Indeed, for each feature $k$, let $t(k)$ be the
  first action that has $k$ as one of its features and image to have
  an urn with balls of two colors, say red and black, and associate an
  extraction from the urn to each action $t\geq t(k)+1$. The initial
  total number of balls in the urn is $c+\sum_{h=1}^{t(k)}W_h$, of
  which $W_{t(k)}$ red.  At each time-step $t\geq t(k)+1$, if the
  extracted ball is red then action $t$ exhibits feature $k$ and the
  composition of the urn is updated with $W_t$ red balls; otherwise,
  action $t$ does not exhibit feature $k$ and the composition of the
  urn is updated with $W_t$ black balls. Therefore quantity
  \eqref{wpa-factor} gives the probability of extracting a red ball at
  time-step $t$.  This is essentially the nodes-features dynamics
  considered in \cite{CRIMALDI2017} with $\delta=0$ only. If we have
  $G_i\leq 1$, an alternative normalizing factor is $B_t=t$. In this
  case the quantity \eqref{wpa-factor} is the empirical mean of the
  random variables $F_{j,k}W_{j}$, with $j=1,\dots, t-1$ (again we
  divide by $t$ instead of $t-1$ for the same reason explained above).

\item[3)] We can extend case 2) to the case in which the fitness
  variables change along time and so we have $W_{t,j,k}=W_{t,j}$
  defined in terms of $G_{t,i(j)}$, where $i(j)$ denotes the agent
  that performs action $j$ and $G_{t, i(j)}$ is her/his fitness at the
  time-step of action $t$, thus following prescription similar to those
  of~\cite{Medo2011, Wang2013}.  We can also extend to the case in
  which the actions can be performed in collaboration by more than one
  agent. In this case the weight $W_{t,j}$ can be defined as a
  function of the fitness at time-step $t$ of all the agents
  performing action $j$.

\item[4)] We can set $W_{t,j,k}=W_{t,k}$ for all $t,j,k$ with $B_t=t$
  so that the term \eqref{wpa-factor} becomes the average popularity
  of feature $k$ ``adjusted'' by the quantity $W_{t,k}$. For instance,
  we can take $W_{t,k}$ as a decreasing function of $t^*(k)=\max\{j :
  1 \leq j \leq t-1\; \hbox{and}\; F_{j,k} = 1\}$, which is the last
  action, before action $t$, that has $k$ among its features. By doing
  so, in \eqref{wpa-factor} the average popularity of $k$ is
  discounted by the lenght of time between the last appearence of
  feature $k$ and $t$. Another possibility is to use a weight $W_{t,
    k}$ in order to give more relevance to the features already shown
  by the same agent performing action $t$ in the previous
  actions. More precisely, we can denote by $i(j)$ the agent that
  performs action $j$ and, for each action $t$, we can define
  $W_{t,k}$ as an increasing function of the sum $\sum_{j=1,\dots,
    t-1,\; i(j)=i(t)} F_{j,k}$ so that the more an agent has exhibited
  feature $k$ in her/his own previous actions, the greater the
  probability that also her/his new action will show feature $k$. An
  additional possibility is to eliminate the dependence on $t$
  and consider weights $W_{t,j,k}=W_k$, where $W_k$ can be seen as a
  ``fitness'' random variable associated to feature $k$.

\item[5)] We can modify case 2) by giving a different meaning to
  $G_i$. Indeed, we can associate to each agent $i$ a positive random
  variable $G_i$ in order to describe her/his ``inclination'' of
  adopting the already appeared features. Then we can define the
  weight $W_{t,j,k}$ as $G_{i(t)}$ (or as a function of it), where
  $i(t)$ denotes the agent performing action $t$. In this way, we have
  $W_{t,j,k}=W_t$ for all $t,j,k$, that is the weights only depend on
  the ``inclination'' of the agent performing the action and, if we
  set $B_t=t$ as in case 4), the term \eqref{wpa-factor} becomes the
  average popularity of feature $k$ ``adjusted'' by the quantity
  $W_{t}$.

\item[6)] Finally, we can take $W_{t,j,k}=W_{j,k}$ (i.e. depending on $j$ and
  $k$, but not on $t$) in order to represent the weight given by the
  agent performing action $j$ to feature $k$ exhibited in this
  action. Therefore the total weight of feature $k$ at time-step $t$
  is the total weight given to feature $k$ by the agents who performed
  the previous actions.

\end{itemize}

These are just general examples of possible weights. We refer to the
following applications to real datasets for special cases of the above
examples. It is worth to note that the weights $W_{t,j,k}$ may be not
independent. For example, in case 5) we have exactly the same weight
for all the actions performed by the same agent.

\section{Applications}
\label{applications}

In this section we present some applications of the model to different
real-world networks.  In the first subsection we briefly illustrate
the general methodology used to analyse the datasets (we refer to the
Appendix for further details). The other subsections contain instead
three examples: we first consider two different collaboration
networks, the first one in the area of Automatic Driving and
downloaded from the IEEE database, the second one in the research
field of High Energy Physics and downloaded from the arXiv
repository. In both cases, the agents are the authors, the agents'
actions are the published papers and the features are all $1$-grams
(nouns and adjectives) included in the title or abstract of each
paper. Thus, the considered features identify the main reserch
subjects treated in the papers. For these applications we make use of
weights of the form $W_{t,j}$ (Subsection \ref{delta-weights}, type
3)), that are defined in terms of a fitness variable associated to the
agents who performed previous action $j$, but measured at the
time-step of the current action $t$.  Finally, we present our last
example: we study the quite popular on-line social network of
Instagram, in which the users are the agents, the agents' actions are
the posted photos and, for each media, the features are the hashtags
included in its description. Thus, the considered features identify
the topics the considered posts refer to. For this example, we adopt
weights of the form $W_{t}$ (Subsection \ref{delta-weights}, type 5)),
that solely depend on some quantity related to the agent performing
the current action $t$, in order to adjust the average popularity of
each feature in~\eqref{wpa-factor}.  A more detailed interpretation of
the considered weights is provided in each subsection.

\subsection{General methodology}
For each considered applications, the analysis develops according to
the same outline: 
\begin{itemize}
\item We present and explain the adopted weights. Then we estimate the
  model parameters, following the procedures described in the
  Appendix, Subsections \ref{estimation} and \ref{methodology}.
\item We consider the behavior of the total number $L_t$ of observed
  features along the time-steps $t$ and we compare it with the
  theoretical one of the model. Moreover, we consider the behavior of
  the total number $e(t)$ of edges in the real actions-features matrix
  and we compare it with the mean number $\mu_e(t)$ of edges obtained
  averaging over $R$ simulated actions-features matrices. (See the
  Appendix, Subsection \ref{appendix-asymptotics} for some theoretical
  results regarding the asymptotic behaviors of $L_t$ and $\mu_e(t)$.)
\item We compare the real and simulated matrices by means of the
  indicators $L_T$, $\overline{O}_T$ and $\overline{N}_T$, defined in
  Subsection \ref{methodology} of the Appendix (see
  \eqref{indicatori-confronto}), that respectively refer to the total
  number of features exhibited by all the $T$ observed actions and the
  averaged number of ``old'' and ``new'' features observed for all the
  $T$ actions.
\item We compute the indicators $\overline{m}_1$ and $\overline{m}_2$,
  defined in Subsection \ref{methodology} of the Appendix (see
  \eqref{indicatori-corrispondenze}), both on the real and simulated
  matrices: the former takes into account the fraction of features
  that have been correctly allocated by the model, while the latter
  refers to the relative error committed in the total number of
  observed features.
\item In order to evaluate the relevance of the weights inside the
  dynamics, we simulate it with all the weights equal to $1$ and
  compute the corresponding values of the previous indicators also in
  this case.
\item Finally, we perform a prediction analysis: we estimate the model
  parameters only on a subset of the observed actions, we simulate the
  rest by means of the model and compare the real and simulated
  matrices. The comparison is performed by means of the indicators
  $\overline{m}_1^*$ and $\overline{m}_2^*$, defined in Subsection
  \ref{methodology} of the Appendix (see
  \eqref{indicatori-predizione}), that respectively take into account
  the number of correctly inferred entries and the relative error in
  the overall number of observed features.
\end{itemize}

\subsection{IEEE dataset for Automatic Driving}
\begin{figure}[t]
\center
\includegraphics[scale=0.7]{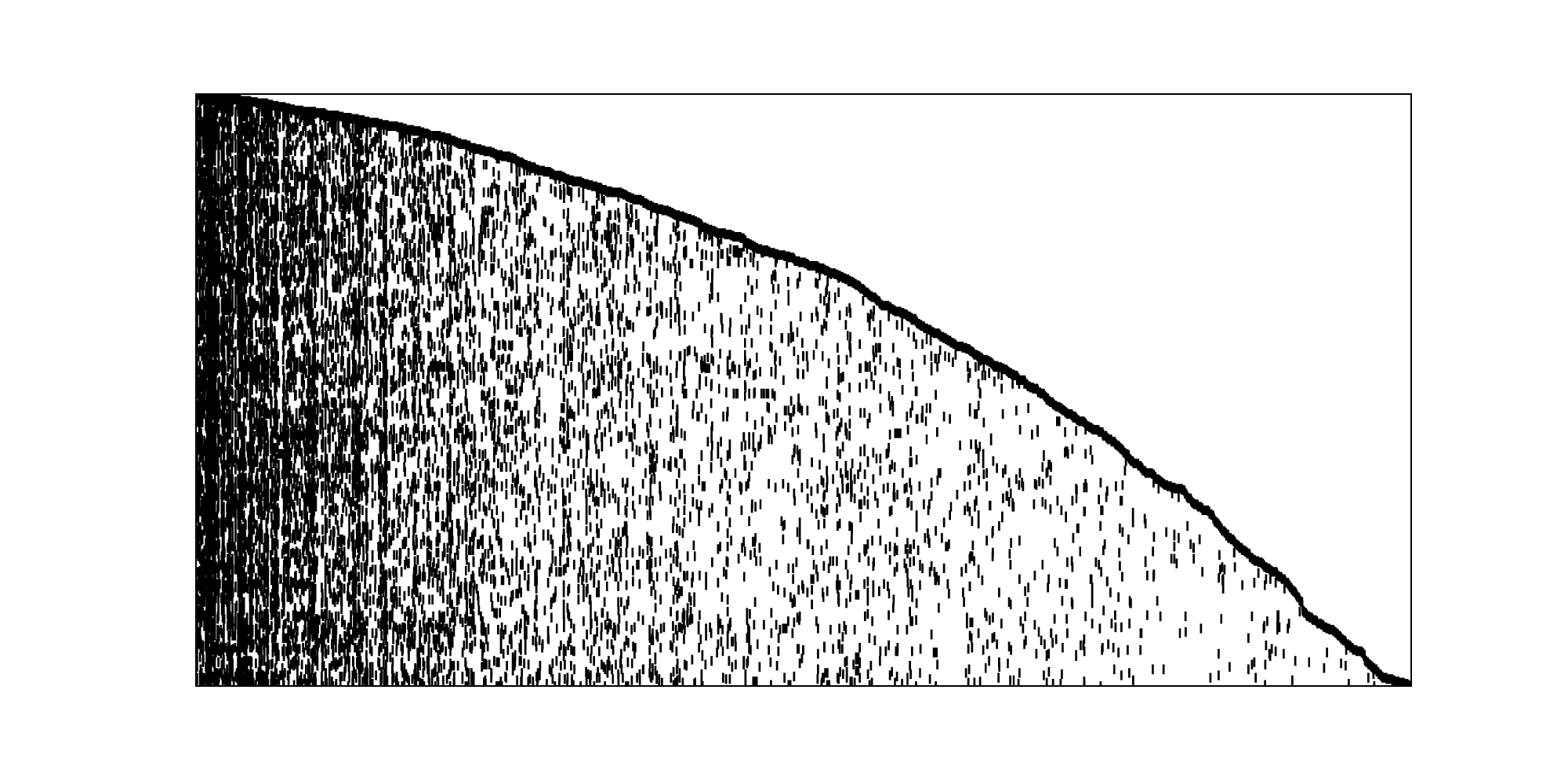}
\caption{\textbf{IEEE Automatic Driving dataset.}  Observed
  actions-features matrix with dimensions $T \times L_T=492\times
  4553$.  Black dots represent $1$ while white dots represent $0$.}
\label{fig:feat_mat_ieee}
\end{figure}

For our first application we have downloaded (on June 26, 2018) all
papers recorded between $2000$ and $2003$ present in the IEEE database
in the scientific research field of Automatic Driving.  We have
performed the research following the same criteria as in
\cite{CRIMALDI2017}, i.e.~selecting all papers containing at least one
of the keywords: Lane Departure Warning, Lane Keeping Assist,
Blindspot Detection, Rear Collision Warning, Front Distance Warning,
Autonomous Emergency Braking, Pedestrian Detection, Traffic Jam
Assist, Adaptive Cruise Control, Automatic Lane Change, Traffic Sign
Recognition, SemiAutonomous Parking, Remote Parking, Driver
Distraction Monitor, V2V or V2I or V2X, Co-Operative Driving,
Telematics \& Vehicles, and Night vision.  The download has yielded
$492$ distinct publications belonging to the required scientific field
and period.  For each paper we have at our disposal all the
bibliographic records, such as title, full abstract, authors' names,
keywords, year of publication, date in which the paper was added to
the IEEE database, and many others. The papers have been sorted
chronologically according to the date in which they were added to the
database. We have considered all nouns and adjectives (from now on
``key-words'') included in the title or abstract as
  the features of our model and sorted them according to their
  ``arrival time''.  (See the Appendix, Subsection \ref{methodology}
  for a more detailed description of the data preparation procedure.)
  The features matrix obtained at the end of the cleaning procedure
  collects $T = 492$ papers (actions) recorded in the period
  $2000-2003$ and involving $n = 1251$ distinct authors (agents) and
  containing $L_T = 4553$ key-words (features). The binary matrix
  entry $F_{t,k}$ indicates whether feature $k$ is present or not into
  the title or the abstract of the paper recorded at time-step $t$. A
pictorial representation of the matrix is provided in Figure
\ref{fig:feat_mat_ieee}. \\

\begin{figure}[t]
\center
\includegraphics[scale=0.4]{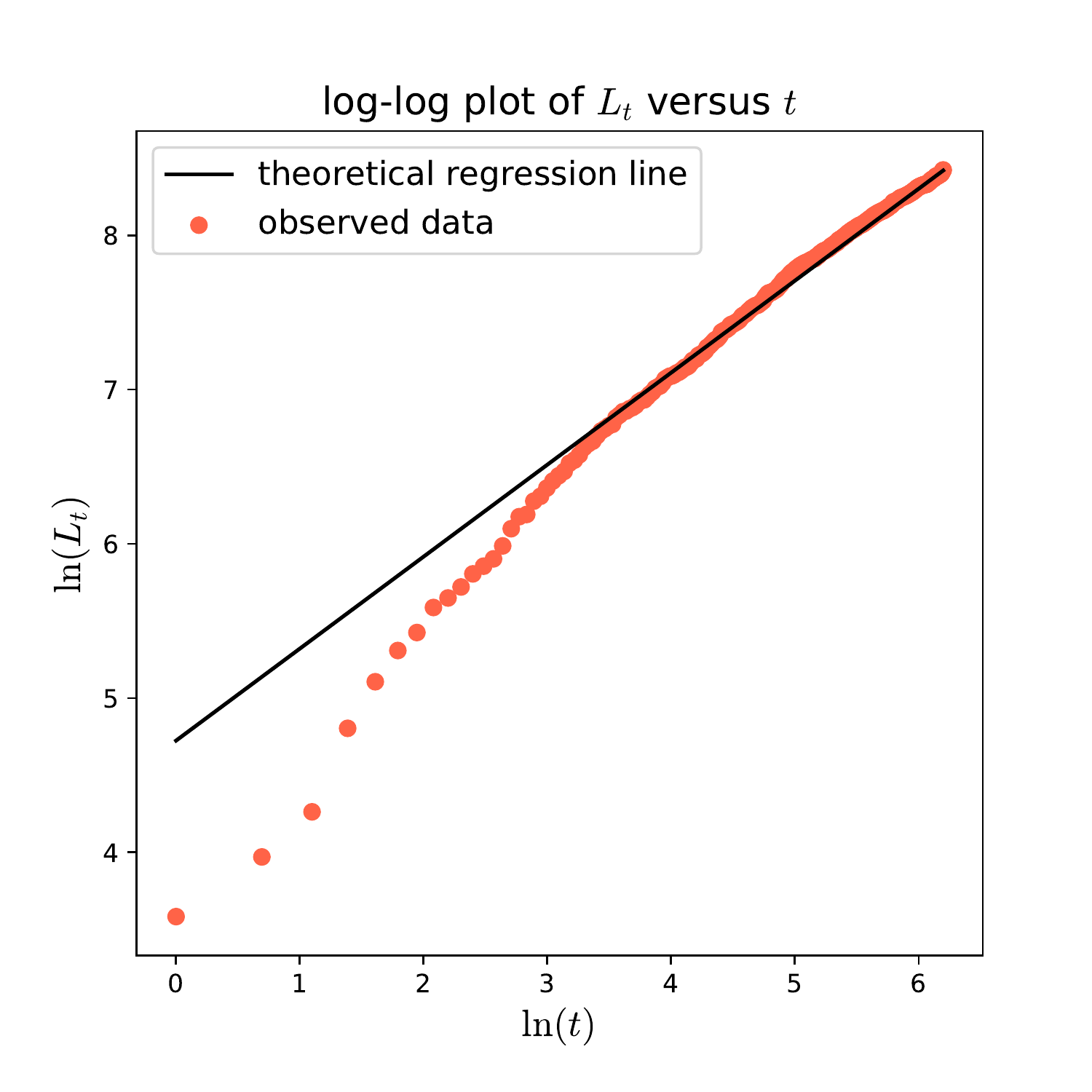}
\includegraphics[scale=0.4]{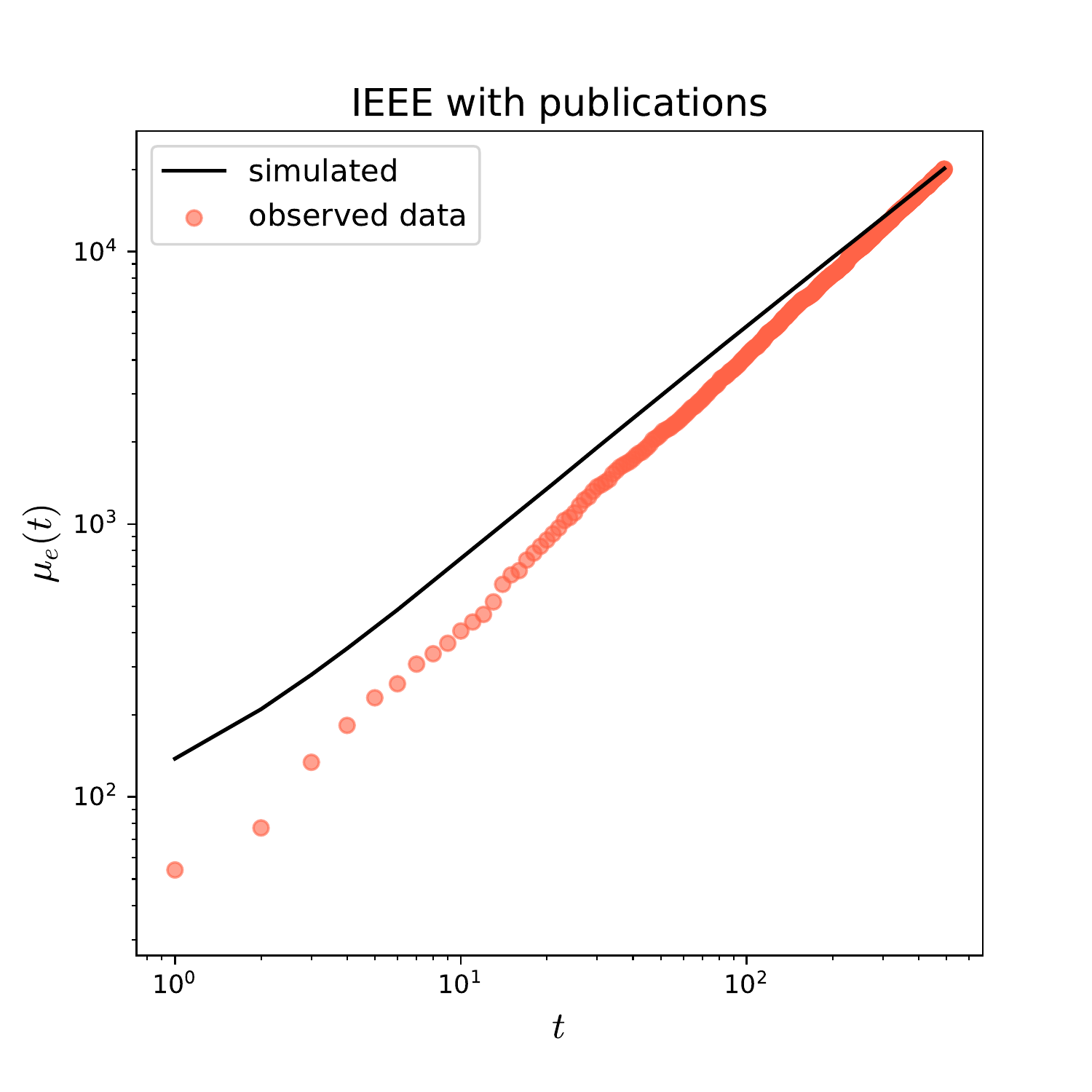}
\caption{\textbf{IEEE Automatic Driving dataset.} Left: Plot of
  $\ln(L_t)$ as a function of $\ln(t)$, with the power-law trend. The
  red dots refer to the real data and the black line gives the
  theoretical regression line with slope $\widehat{\beta}$. Right:
  Asymptotic behavior of the number of edges in the actions-features
  network. Red dots refer to $e(t)$ of the real data, while the black
  line shows $\mu_e(t)$ obtained by the model with $G_{t,i}^{pub}$
  (averaging over $R = 100$ simulations).}
\label{fig:beta_plot_ieee}
\end{figure}

\indent For this application, we use weights of the type 3),
Sec. \ref{delta-weights}. Indeed, at each time-step $t$, we associate
to each author $i$ a ``fitness'' variable $G_{t,i}$ that quantifies the
influence of author $i$ in the considered research field, and we
define the weights as
\begin{equation}\label{eq:weights_physics}
\begin{split}
W_{t,j,k} &= W_{t,j} = e^{-1/M_{t,j}} \; \hbox{ with } \; 
M_{t,j} = \max\{G_{t,i}:\, i\in \mathcal{I}(j)\} \; \hbox{where} \;\\
\mathcal{I}(j)&=\hbox{set of the agents performing action $j$}\,.
\end{split}
\end{equation}
Therefore the inclusion probability in equation \eqref{incl_prob} reads as
\begin{equation}\label{eq:inc_prob_physics}
P_t(k) = \frac{\delta}{2} + 
(1-\delta) \frac{\sum_{j = 1}^{t-1} F_{j,k}\,e^{-1/M_{t,j}}}{t}.
\end{equation}
\noindent The term $M_{t,j}$ is the maximum among the
  fitness variables $G_{t,i}$ at time-step $t$ of all the authors
  $i\in{\mathcal I}(j)$, i.e. the authors who published the paper
  appeared at time-step $j$.  A high value of $G_{t,i}$ should
  identify a person who is relevant in the considered research field
  so that it is likely that other scholars use the same features of
  her/his actions, that essentially are the keywords related to
  her/his research.
As a consequence, in the preferential attachment term, we give to each
old feature $k$ a weight that is increasing with respect to the
fitness variables of the authors who included $k$ in their papers. We
analyse two different fitness variables:
\begin{equation}\label{fitness-pub}
G_{t,i}^{pub}=\hbox{(total number of author $i$'s publications until
  time-step $t-1$)} + 1
\end{equation}
and 
\begin{equation}\label{fitness-col}
G_{t,i}^{col}=\hbox{(total number of author $i$'s collaborators until
  time-step $t-1$)} + 1.
\end{equation}
(Note that we count publications or collaborators until time-step
$t-1$, that is until a time-step before the record time of paper
$t$ and $1$ is added in order to avoid division by zero in the
previous formula \eqref{eq:weights_physics}.)
\\

\begin{table}[tp]
\centering
\begin{tabular}{lcccc}
\toprule
$p$ & $\widehat{p}$ & $\overline{p}$ & $MSE(p)$\\
\midrule
$\alpha$ & $68.534$ & $68.699$ & $14.699$ \\
$\beta$ & $0.5962$ & $0.5962$ & $0.0001$\\
$\delta$ with $G_{t,i}^{pub}$ & $\approx 2.22 \cdot 10^{-16}$ & $\approx 4.36 \cdot 10^{-5}$ & $\approx 5.90 \cdot 10^{-9}$\\
$\delta$ with $G_{t,i}^{col}$ & $\approx 2.26 \cdot 10^{-16}$ & $\approx 5.01 \cdot 10^{-5}$ & $\approx 6.82 \cdot 10^{-9}$\\
\bottomrule
\end{tabular}
\caption{\textbf{IEEE Automatic Driving dataset.} Estimation of the
  model parameters. The average values $\overline{p}$ and the parameters' 
  $MSE$ are computed over $R = 100$ realizations of the model. See Subsection 
  \ref{methodology} for further details.}
\label{table:estimation_ieee}
\end{table}

\begin{table}[tp]
\centering
\begin{tabular}{lccccccccccc}
\toprule
Matrix & $L_T$ & $\sigma_{L_T}$ & $\overline{O}_T$ & $\sigma_{O_T}$ & $\overline{N}_T$ & $\sigma_{N_T}$\\
\midrule
real & $4553$ & & $31.54$ & & $9.25$ \\
Weights with $G_{t,i}^{pub}$ & $4558$ & $68.19$ & $\mathbf{32.13}$ & $1.29$ & $9.26$ & $0.14$\\
Weights with $G_{t,i}^{col}$ & $4548$ & $63.00$ & $54.20$ & $2.88$ & $9.25$ & $0.13$ \\
Weights = 1 & $4551$ & $71.50$ & $134.73$ & $3.48$ & $9.25$ & $0.15$ \\
\bottomrule
\end{tabular}
\caption{\textbf{IEEE Automatic Driving dataset.} Comparison between
  real and simulated actions-features matrices by means of the
  indicators \eqref{indicatori-confronto}. For the simulations, all
  the considered quantities have been averaged over $R = 100$
  realizations of the model. We also include an estimate of the
  variations around the averaged values, through the computation of
  the sample standard deviations. See Subsection \ref{methodology} for
  further details.}
\label{table:comparison_ieee}
\end{table}

\begin{table}[tp]
\centering
\begin{tabular}{lcc}
\toprule
Weights with $G_{t,i}^{pub}$ & $\overline{m}_1$ & $\overline{m}_2$ \\
\midrule
$k^* = 4553$ (all observed features) & $\mathbf{0.97}$ & $0.047$ \\
$k^* = 100$ & $\mathbf{0.88}$ & \\
$k^* = 200$ & $\mathbf{0.90}$ & \\ 
$k^* = 300$ & $\mathbf{0.91}$ & \\
\bottomrule
\end{tabular}
\\
\begin{tabular}{lcc}
\toprule
Weights with $G_{t,i}^{col}$ & $\overline{m}_1$ & $\overline{m}_2$ \\
\midrule
$k^* = 4553$ (all observed features) & $0.96$ & $0.049$ \\
$k^* = 100$ & $0.83$ & \\
$k^* = 200$ & $0.86$ & \\ 
$k^* = 300$ & $0.88$ & \\
\bottomrule
\end{tabular}
\\
\begin{tabular}{lcc}
\toprule
Weights = 1 & $\overline{m}_1$ & $\overline{m}_2$ \\
\midrule
$k^* = 4553$ (all observed features) & $0.93$ & $0.049$ \\
$k^* = 100$ & $0.55$ & \\
$k^* = 200$ & $0.64$ & \\      
$k^* = 300$ & $0.70$ & \\
\bottomrule
\end{tabular}
\caption{\textbf{IEEE Automatic Driving dataset.} Comparison between
  real and simulated actions-features matrices by means of the
  indicators \eqref{indicatori-corrispondenze}. The first row of each
  table evaluates the indicators on the whole matrix ($k^*=4553$),
  while the other rows show the results computing the indicator
  $\overline{m}_1$ only on the first $k^*$ ($=100,\,200,\,300$)
  features. See Subsection \ref{methodology} for further details.}
\label{table:prediction_ieee_2}
\end{table}

\begin{table}[tp]
\centering
\begin{tabular}{lcc}
\toprule
Weights with $G_{t,i}^{pub}$ & $\overline{m}_1^*$ & $\overline{m}_2^*$ \\
\midrule
$T^* = 369$ and $k^* = 4553$ & $0.99$ & $0.017$ \\
$T^* = 246$ and $k^* = 4553$ & $0.98$ & $0.060$ \\
$T^* = 123$ and $k^* = 4553$ & $0.98$ & $0.113$ \\
\midrule
$T^* = 369$ and $k^* = 200$ & $0.93$ & \\
$T^* = 246$ and $k^* = 200$ & $0.93$ & \\
$T^* = 123$ and $k^* = 200$ & $0.93$ & \\
\bottomrule
\end{tabular}
\caption{\textbf{IEEE Automatic Driving dataset.} Predictions on the
  actions-features matrix. The indicators
  \eqref{indicatori-predizione} are computed for different levels of
  information used as ``training set'': more precisely, the different
  $T^*$ correspond to $75\%$, $50\%$ and $25\%$ of the set of the
  actions, respectively. Moreover, the indicator $\overline{m}_1^*$ is
  computed on the whole matrix ($k^*=4553$) and also taking into
  account only the first $k^*=200$ features. See Subsection
  \ref{methodology} for further details.}
\label{table:prediction_papers_ieee}
\end{table}

\indent For both the definitions of fitness, we perform the analysis
following the methodology explained in Subsection \ref{methodology}
(for the simulated matrices, all the considered quantities have been
averaged over $R = 100$ realizations of the model).  We first estimate
the model's parameters, obtaining the results in Table
\ref{table:estimation_ieee}. We can see that the weighted preferential
attachment term \eqref{wpa-factor} plays a predominant role, due to
the estimated value obtained for the parameter $\delta$ that is
approximately zero. Figure \ref{fig:beta_plot_ieee} provides in the
left panel a log-log plot of the cumulative count of new features
(key-words) as a function of time (see the red dots), that clearly
shows a power-law behavior. This agrees with the theoretical property
of the model stated in the Appendix, Subsec.~\ref{total-num-features},
according to which the power-law exponent has to be equal to the
parameter $\beta$.  This fact is checked in the plot by the black
line, whose slope is the estimated value of the parameter $\beta$.
The goodness of fit of our model to the dataset has been evaluated
through the computation of the quantities \eqref{indicatori-confronto}
and \eqref{indicatori-corrispondenze}.  These results are shown in
Table \ref{table:comparison_ieee} and Table
\ref{table:prediction_ieee_2}. Table \ref{table:comparison_ieee} shows
that our model reproduces the total number $L_T$ of features observed
at the end of the observation period $T$, as well as the average
number of new features $\overline{N}_T$ in all the three considered
cases.  The average number of old features (i.e.~the quantity
$\overline{O}_T$) is well reproduced only in the case with
$G_{t,i}^{pub}$ (that is the case with the fitness based on the number
of publications). Table \ref{table:prediction_ieee_2} also indicates
that the model with $G_{t,i}^{pub}$, or with $G_{t,i}^{col}$, shows a
better performance than the one with all the weights equal to one.
More precisely, the values obtained for the indicator $\overline{m}_2$
are almost the same for all the three cases (the average error on the
total number of arrived features is around $5\%$); while the most
significant differences are in the values of the indicator
$\overline{m}_1$. Indeed, for the model with the fitness
$G_{t,i}^{pub}$, the computed value of $\overline{m}_1$ ranges from
$88\%$ to $97\%$, pointing out that a high percentage of the entries
in the actions-features matrix have been correctly inferred by the
model. The same value for the model with the fitness $G_{t,i}^{col}$
ranges from $83\%$ to $96\%$, and, for the model with all the weights
equal to $1$, it ranges from $55\%$ to $93\%$. The differences with
respect to the case with all the weights equal to $1$ are more evident
when we select the first $k^*$ features: indeed, with $G_{t,i}^{pub}$
we succeed to infer the value of at least $88\%$ of the entries; while
with all the weights equal to $1$ the percentage remains under
$70\%$. This means that the major difference in the performance of the
different considered weights is in the first features, that are those
for which the preferential attachment term is more relevant. At this
point, we choose the model that takes into account the authors' number
of publications as the best performing one for the considered dataset
and in the following we focus on it. In Table
\ref{table:prediction_papers_ieee} we evaluate the predictive power of
the model: we estimate the parameters of the model only on a subset of
the observed actions, respectively the $75\%$, $50\%$ and $25\%$ of
the total observations; we then predict the features for the
``future'' actions $\{T^*+1, \dots, T\}$ and compare the predicted and
observed results by means of the indicators in
\eqref{indicatori-predizione} over the whole set of features and only
on a portion of it. The indicator $\overline{m}_1^*$ ranges from
$93\%$ to $99\%$. Finally, in the right panel of Figure
\ref{fig:beta_plot_ieee}, we provide the asymptotic behavior of the
number of edges in the actions-features network: more precisely, the
red dots represent the total number $e(t)$ of edges observed in the
real actions-features matrix at each time-step; while the continuous
black line shows the mean number $\mu_e(t)$ of edges obtained
averaging over $R = 100$ simulations of the model with the chosen
weights.  \\

\indent It is worth to note that the difference in the performance
between the two definitions of fitness variables has a straightforward
interpretation: in the considered case, i.e. for the publications in
the area of Automatic Driving in the considered period,
the relevance of an author (with respect to the
  probability of transmitting her/his features) is better measured by
  considering the number of her/his publications rather than the
  number of her/his co-authors. As we will see later on, we get a
  different result for our second application.

\subsection{ArXiv dataset for Theoretical High Energy Physics}

\begin{figure}[tp]
\centering
\includegraphics[scale=0.7]{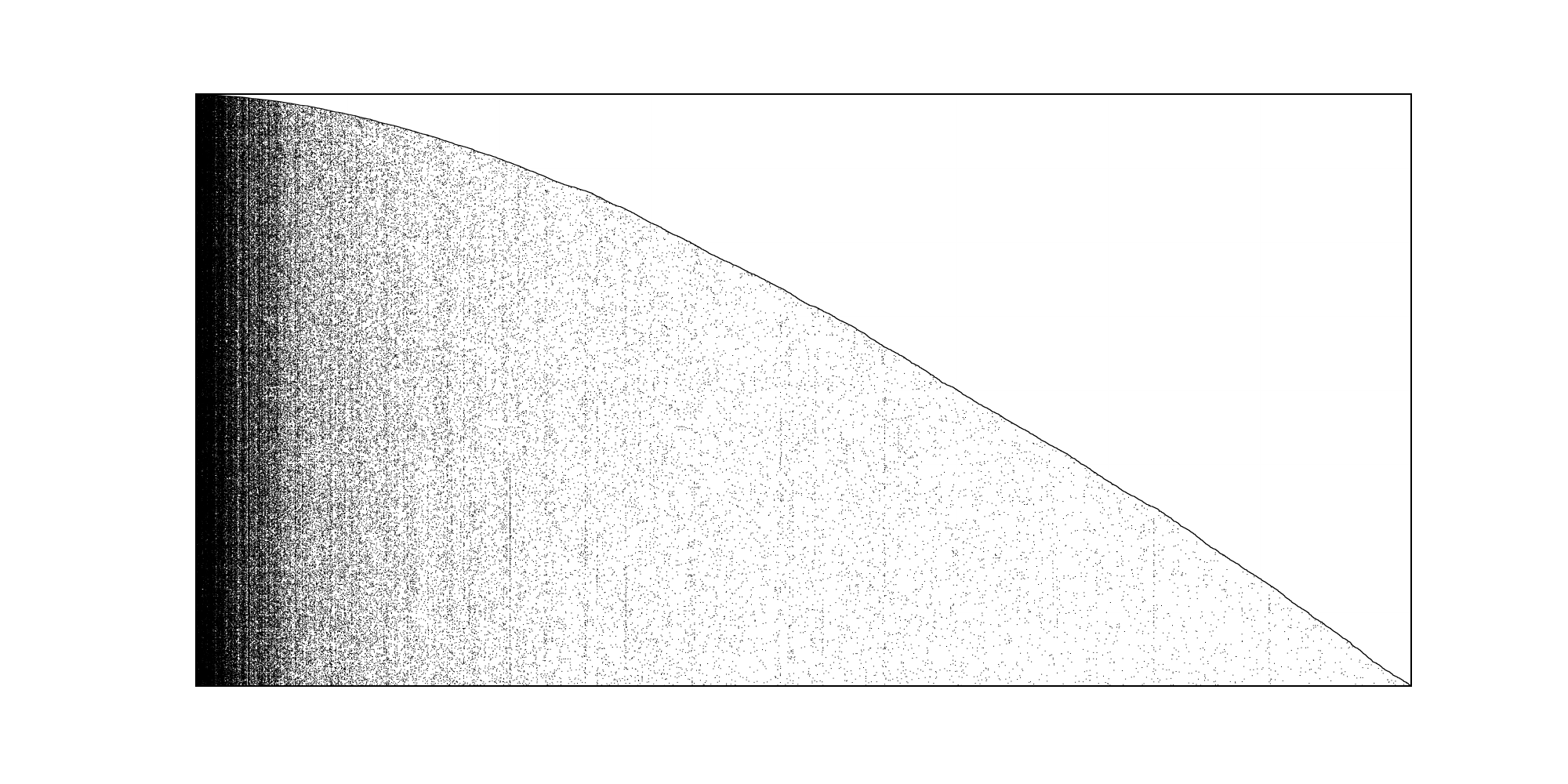}
\caption{\textbf{arXiv High Energy Physics dataset.}  Observed
  actions-features matrix with dimensions $T \times L_T=10603\times
  22304$.  Black dots represent $1$ while white dots represent $0$.}
\label{fig:feat_mat_coauthor}
\end{figure}

Our second application has been performed with the arXiv dataset of
publications in the scientific area of Theoretical High Energy Physics
(Hep-Th), recorded in the period $2000-2003$, freely available from
\cite{Leskovec2006}\footnote{
  \url{https://snap.stanford.edu/data/ca-HepTh.html}}.  The dataset
collects a sample of text files reporting the full frontispiece of
each paper, so we have information on: arXiv id number, date of
submission, name and email of the author who made the submission,
title, authors' names and the entire text of the abstract. From the
original format we isolate the submission date and the identity number
of the paper, in order to sort all papers (actions)
chronologically. Then, with the final purpose of constructing the
features matrix, we consider all key-words
included either in the main title or in
the abstract as the features of the papers and we sort them according
to their time of appearence. (The complete data preparation phase is
described in the Appendix, Subsection \ref{methodology}).  We
constructed the features matrix $F$, whose elements
  are equal to $F_{t,k} = 1$ if paper $t$ includes word $k$ either
in the title or in the abstract and $F_{t,k} = 0$ otherwise. The
result is shown in Figure \ref{fig:feat_mat_coauthor}, where the
observed actions-features matrix collects $T = 10603$ papers (actions)
registered between $2000$ and $2003$ and $L_T = 22304$ key-words
appeared in the title or in the abstract (features),
while the total number of involved authors (agents) is $n = 5633$. \\

\begin{figure}[tp]
\center
\includegraphics[scale=0.4]{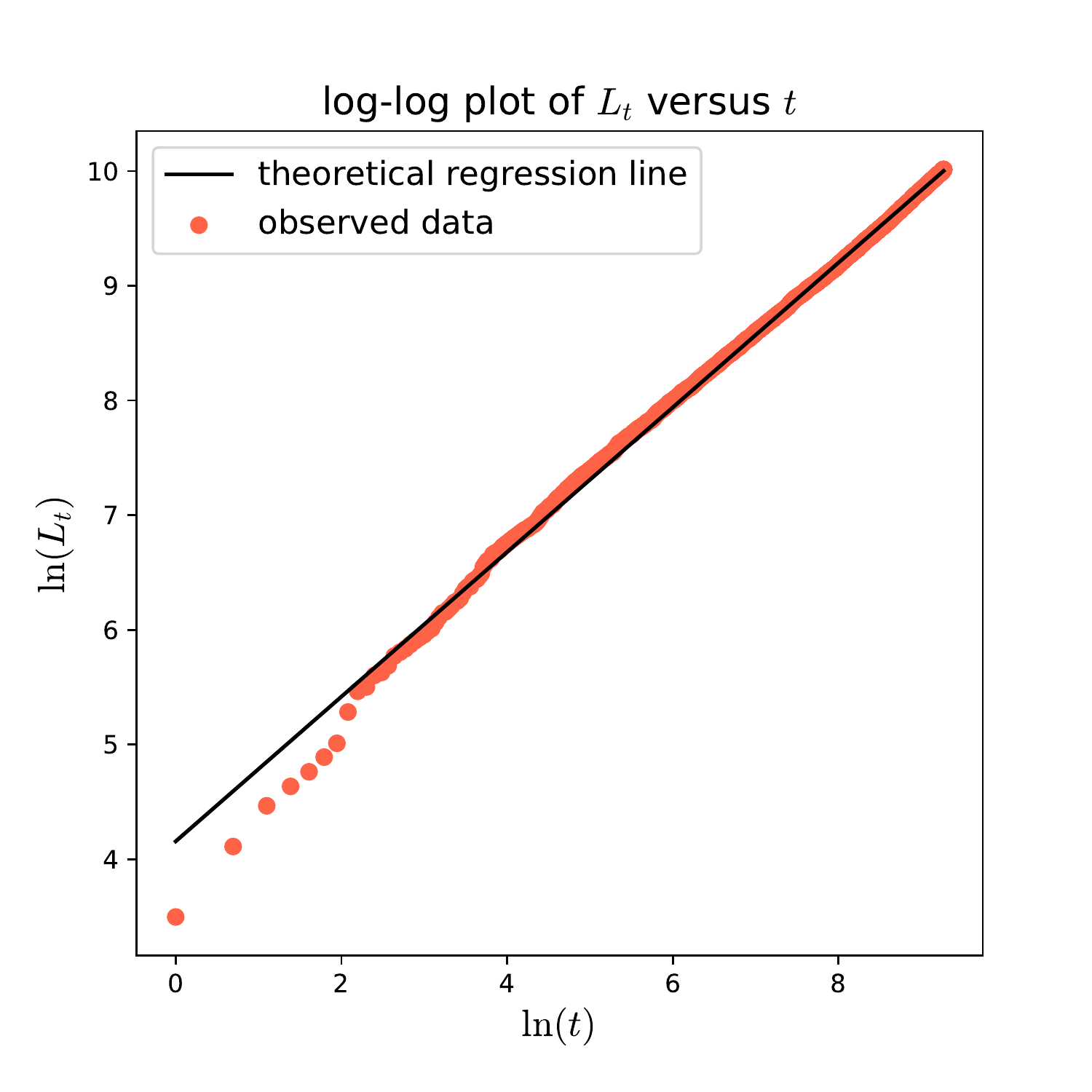}
\includegraphics[scale=0.4]{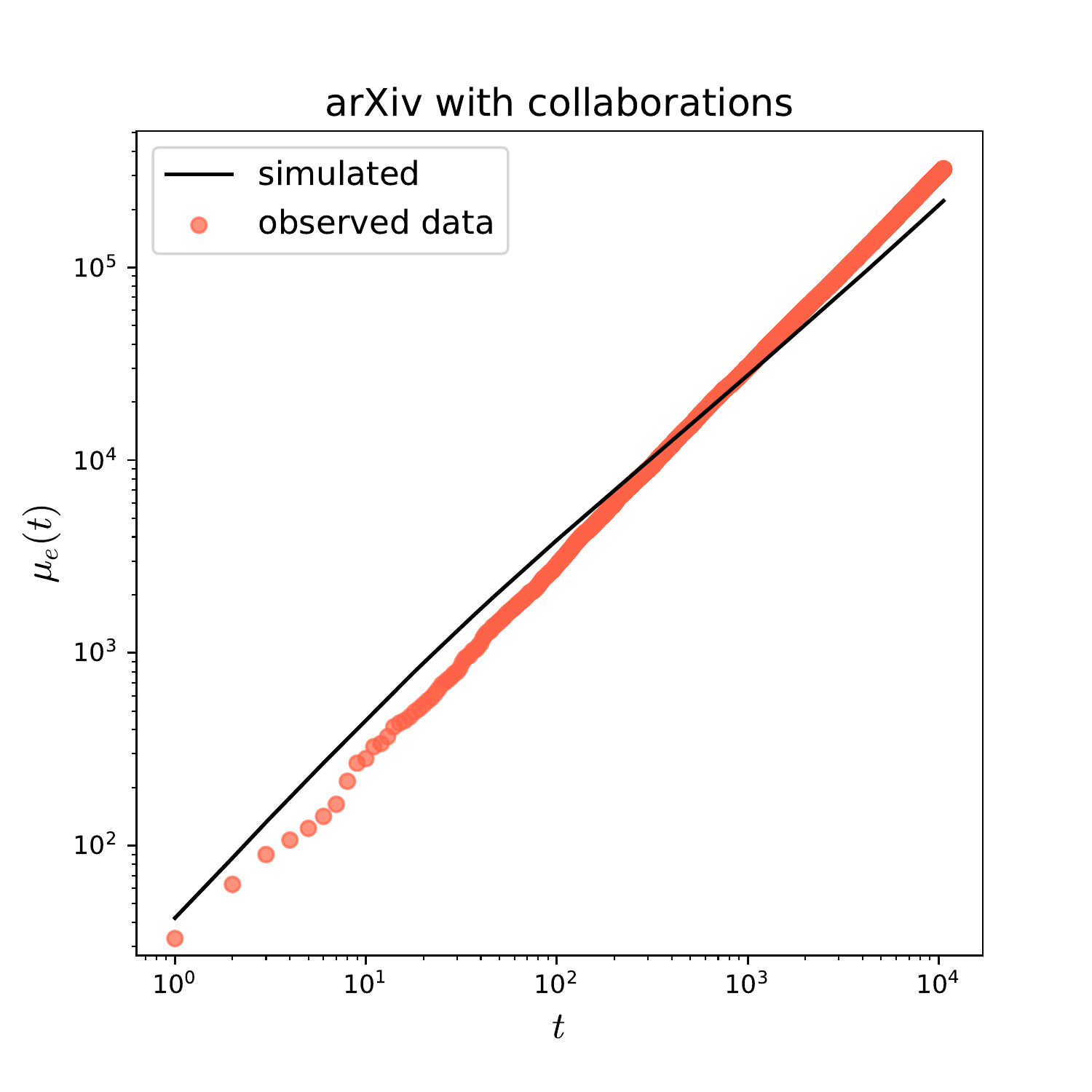}
\caption{\textbf{arXiv High Energy Physics dataset.}  Left: Plot of
  $\ln(L_t)$ as a function of $\ln(t)$, with the power-law trend. The
  red dots refer to the real data and the black line gives the
  theoretical regression line with slope $\widehat{\beta}$. Right:
  Asymptotic behavior of the number of edges in the actions-features
  network. Red dots refer to $e(t)$ of the real data, while the black
  line shows $\mu_e(t)$ obtained by the model with $G_{t,i}^{col}$
  (averaging over $R = 100$ simulations).}
\label{fig:beta_plot_physics}
\end{figure}

\begin{table}[tp]
\centering
\begin{tabular}{lcccc}
\toprule
$p$ & $\widehat{p}$ & $\overline{p}$ & $MSE(p)$\\
\midrule
$\alpha$ & $40.812$ & $40.940$ & $2.4980$ \\
$\beta$ & $0.6305$ & $0.6303$ & $2.35 \cdot 10^{-5}$\\
$\delta$ with $G_{t,i}^{pub}$ & $\approx 2.22 \cdot 10^{-16}$ & $\approx 1.41 \cdot 10^{-6}$ & $\approx 5.55 \cdot 10^{-12}$ \\
$\delta$ with $G_{t,i}^{col}$ & $\approx 2.22 \cdot 10^{-16}$ & $\approx 1.06 \cdot 10^{-6}$ & $\approx 3.44 \cdot 10^{-12}$ \\
\bottomrule
\end{tabular}
\caption{\textbf{arXiv High Energy Physics dataset.} 
Estimation of the model parameters. The average values $\overline{p}$ 
and the parameters' $MSE$ are computed over $R=100$ realizations of the 
model. See Subsection \ref{methodology} for further details.}
\label{table:estimation_arxiv}
\end{table}

\begin{table}[tp]
\centering
\begin{tabular}{lccccccc}
\toprule
Matrix & $L_T$ & $\sigma_{L_T}$ & $\overline{O}_T$ & $\sigma_{O_T}$ & $\overline{N}_T$ & $\sigma_{N_T}$ \\
\midrule
real & $22304$ & & $28.42$ & & $2.10$ & \\
Weights with $G_{t,i}^{pub}$ & $22305$ & $145.83$ & $15.37$ & $0.41$ & $2.10$ & $0.02$ \\
Weights with $G_{t,i}^{col}$ & $22317$ & $148.35$ & $18.79$ & $0.70$ & $2.10$ & $0.02$ \\
Weights = 1 & $22313$ & $147.35$ & $97.16$ & $5.16$ & $2.10$ & $0.02$ \\
\bottomrule
\end{tabular}
\caption{\textbf{arXiv High Energy Physics dataset.} Comparison
  between real and simulated actions-features matrices by means of the
  indicators \eqref{indicatori-confronto}. For the simulations, all
  the considered quantities have been averaged over $R = 100$
  realizations of the model. We also include an estimate of the
  variations around the averaged values, through the computation of
  the sample standard deviations. See Subsection \ref{methodology} for
  further details.}
\label{table:comparison_arxiv}
\end{table}

\begin{table}[tp]
\centering
\begin{tabular}{lcc}
\toprule
Weights with $G_{t,i}^{pub}$ & $\overline{m}_1$ & $\overline{m}_2$ \\
\midrule
$k^* = 22304$ (all observed features) & $0.995$ & $0.022$ \\
$k^* = 5576$ & $0.992$ & \\
$k^* = 11152$ & $0.994$ & \\ 
$k^* = 16728$ & $0.995$ & \\
\bottomrule
\end{tabular}\\
\begin{tabular}{lcc}
\toprule
Weights with $G_{t,i}^{col}$ & $\overline{m}_1$ & $\overline{m}_2$ \\
\midrule
$k^* = 22304$ (all observed features) & $0.995$ & $0.021$ \\
$k^* = 5576$ & $0.991$ & \\
$k^* = 11152$ & $0.994$ & \\ 
$k^* = 16728$ & $0.994$ & \\
\bottomrule
\end{tabular}\\
\begin{tabular}{lcc}
\toprule
Weights = 1 & $\overline{m}_1$ & $\overline{m}_2$ \\
\midrule
$k^* = 22304$ (all observed features) & $0.987$ & $0.021$ \\
$k^* = 5576$ & $0.976$ & \\
$k^* = 11152$ & $0.985$ & \\      
$k^* = 16728$ & $0.986$ & \\
\bottomrule
\end{tabular}
\caption{\textbf{arXiv High Energy Physics dataset.} Comparison
  between real and simulated actions-features matrices by means of the
  indicators \eqref{indicatori-corrispondenze}.  The first row of each
  table evaluates the indicators on the whole matrix ($k^*=22304$),
  while the other rows show the results computing the indicator
  $\overline{m}_1$ only on the first $k^*$ ($=5576,\,11152,\,16728$)
  features. See Subsection \ref{methodology} for further details.}
\label{table:prediction_papers_2}
\end{table}

\indent The weights for this application are defined as in the
previous one, described in equation \eqref{eq:weights_physics}.  We
consider again the two different definitions for the ``fitness'' term
$G_{t,i}$ (see \eqref{fitness-pub} and \eqref{fitness-col}).  The
performed analysis follows the outline explained in Subsection
\ref{methodology} (for the simulated matrices, all the considered
quantities have been averaged over $R = 100$ realizations of the
model).  We first estimate the model's parameters, obtaining the
results in Table \ref{table:estimation_arxiv}. We see that the
weighted preferential attachment term \eqref{wpa-factor} gives most of
the contribution due to the estimated value obtained for the parameter
$\delta$ that is essentially zero.  Figure \ref{fig:beta_plot_physics}
provides in the left panel a log-log plot of the cumulative count of
new features (key-words) as a function of time (see the red dots),
that clearly shows a power-law behavior. This agrees with the
theoretical property of the model stated in the Appendix,
Subsec.~\ref{total-num-features}, according to which the power-law
exponent has to be equal to the parameter $\beta$.  This fact is
checked in the plot by the black line, which slope is the estimated
value of the parameter $\beta$. The goodness of fit of our model to
the dataset has been evaluated through the computation of the
quantities \eqref{indicatori-confronto} and
\eqref{indicatori-corrispondenze}.  These results are shown in Table
\ref{table:comparison_arxiv} and Table
\ref{table:prediction_papers_2}. Table
  \ref{table:comparison_arxiv} shows that our model is able to
  reproduce the total number $L_T$ of features observed at the end of
  the observation period $T$ and the average number of new features
  $\overline{N}_T$. Instead, the average number of old features
  (i.e.~the quantity $\overline{O}_T$) is under-estimated by the model
  with the weights based on $G_{t,i}^{pub}$ and $G_{t,i}^{col}$, while
  it is widely over-estimated in the case with all the weights equal
  to $1$. The discrepancy in the values is smaller for the
  case with $G_{t,i}^{col}$ (that is the case with the fitness based
  on the number of collaborators).
Table \ref{table:prediction_papers_2} shows that the performance of 
the model in reproducing the data are comparable, with both the 
considered definitions of fitness and they are also good for the 
case with all weights equal to one.  
At this point, we choose the model that takes into account the
authors' number of collaborations and the last analysis focuses on
it. In Table \ref{table:prediction_papers} we evaluate the predictive
power of the model: we estimate the parameters of the model only on a
subset of the observed actions, respectively the $75\%$, $50\%$ and
$25\%$ of the total observations; we then predict the features for the
``future'' actions $\{T^*+1, \dots, T\}$ and compare the predicted and
observed results by means of the indicators in
\eqref{indicatori-predizione} over the whole set of features and only
on a portion of it. In particular, we obtain that the
  indicator $\overline{m}_1^*$ is almost always equal to $99\%$. 
Finally, in the right panel of Figure \ref{fig:beta_plot_physics}, we
provide the asymptotic behavior of the number of edges in the
actions-features network: more precisely, the red dots represent the
total number $e(t)$ of edges observed in the real actions-features
matrix at each time-step; while the continuous black line shows the
mean number $\mu_e(t)$ of edges obtained averaging over $R = 100$
simulations of the model with the chosen weights.  \\

\begin{table}[t]
\centering
\begin{tabular}{lccc}
\toprule
Weights with $G_{t,i}^{col}$ & $\overline{m}_1^*$ & $\overline{m}_2^*$ \\
\midrule
$T^* = 7952$ and $k^* = 22304$ & $0.997$ & $0.006$ \\
$T^* = 5302$ and $k^* = 22304$ & $0.997$ & $0.026$ \\
$T^* = 2651$ and $k^* = 22304$ & $0.996$ & $0.035$ \\
\midrule
$T^* = 7952$ and $k^* = 11152$ & $0.996$  & \\
$T^* = 5302$ and $k^* = 11152$ & $0.996$  & \\
$T^* = 2651$ and $k^* = 11152$ & $0.996$  & \\
\bottomrule
\end{tabular}\\
\caption{\textbf{arXiv High Energy Physics dataset.} Predictions on
  the actions-features matrix. The indicators
  \eqref{indicatori-predizione} are computed for different levels of
  information used as ``training set'': more precisely, the different
  $T^*$ correspond to $75\%$, $50\%$ and $25\%$ of the set of the
  actions, respectively. Moreover, the indicator $\overline{m}_1^*$ is
  computed on the whole matrix ($k^*=22304$) and also taking into
  account only the first $k^*=11152$ features. See Subsection 
  \ref{methodology} for further details.}
\label{table:prediction_papers}
\end{table}

\indent Contrarily to the previous case, in this application we
observe a comparable performance of the model with
  both the considered definitions of fitness. This means that, for
the publications in High Energy Physics in the considered period,
both the number of co-authors and the number of
  publications of an author can be considered as reasonable measures
  in order to evaluate her/his relevance in the research field.

\subsection{Instagram dataset}
\begin{figure}[tp]
\center
\includegraphics[scale=0.7]{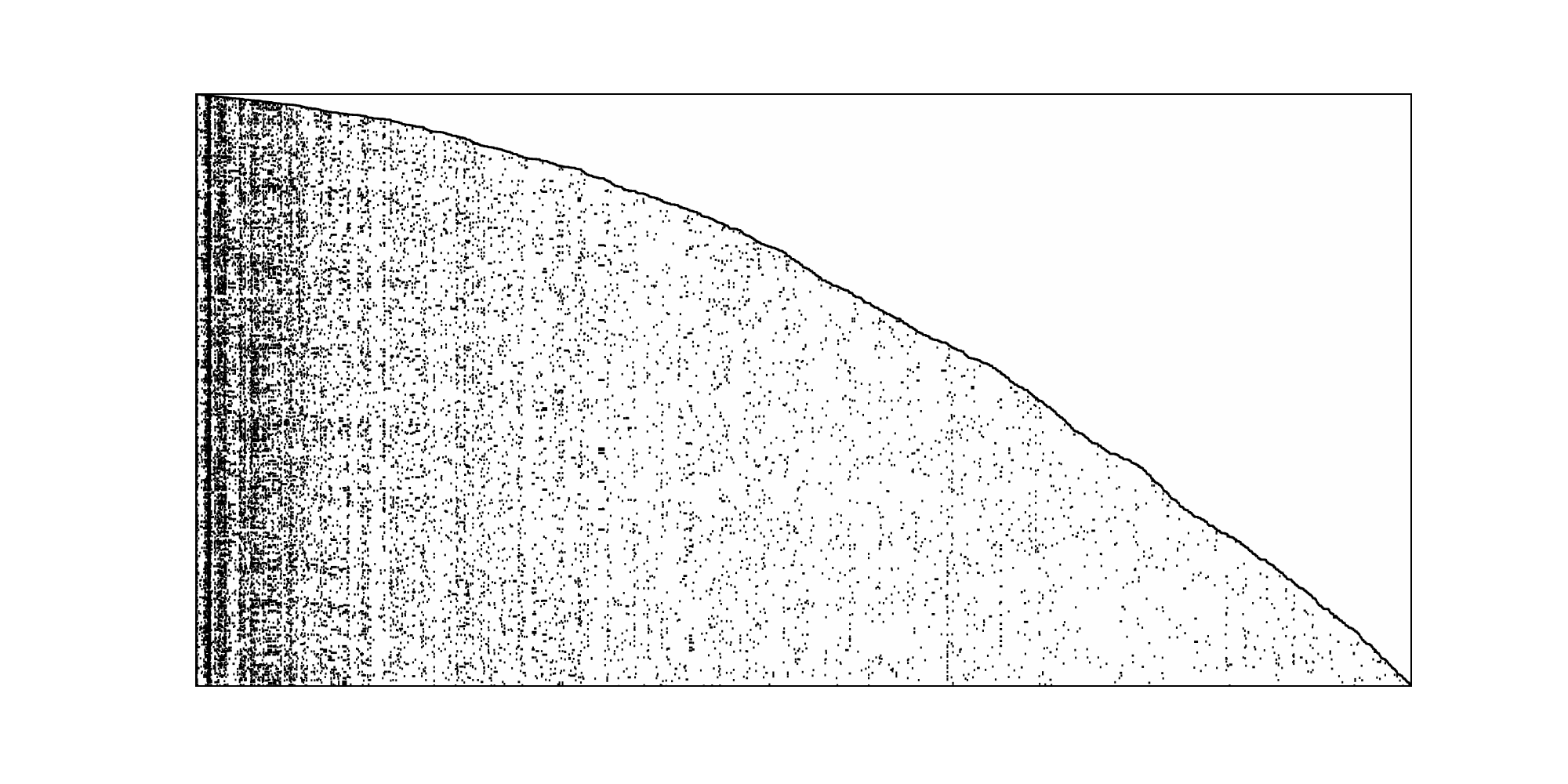}
\caption{\textbf{Instagram dataset.} Observed actions-features matrix,
  with dimesions $T \times L_T=2151\times 5890$.  Black dots represent
  $1$ while white dots represent $0$.}
\label{real_simuls_feat}
\end{figure}

The dataset used for this application has been kindly provided
by~Prof. Emilio
Ferrara\footnote{\url{http://www.emilio.ferrara.name/datasets/}} and a
more detailed description can be found in \cite{Ferrara2014}. The
dataset has been crawled through the Instagram API between January 20
and February 17, 2014 and collects public media (with their author,
timestamp and set of hashtags) as well as users information (with
their list of followers and followees) of a set of $2100$ anonymized
partecipants to $72$ popular photographic contests that took place
between October $2010$ and February $2014$. The overall media dataset
records more than one million posts but, with the purpose of
maximizing the density of our actions-features matrix, we considered
only those posts posted during the weekends in the crawling period
(Jan 20$-$Feb 17, 2014) in which at least $5$ hashtags are used.  This
procedure yields a sample of $T = 2151$ posts (actions) and $L_T =
5890$ hashtags (features). The available posts were ordered
chronologically according to the associate timestamp of publication
and the hashtags (features) were sorted in terms of their first
appearence in a post. After this first phase of data arrangement, we
constructed the actions-features matrix $F$, with $F_{t,k} = 1$ if
post $t$ contains hashtag $k$ and $F_{t,k} = 0$ otherwise. The
resulting matrix is shown in Figure \ref{real_simuls_feat}, with
non-zero values indicated by black points.  \\

\begin{figure}[tp]
\center
\includegraphics[scale=0.4]{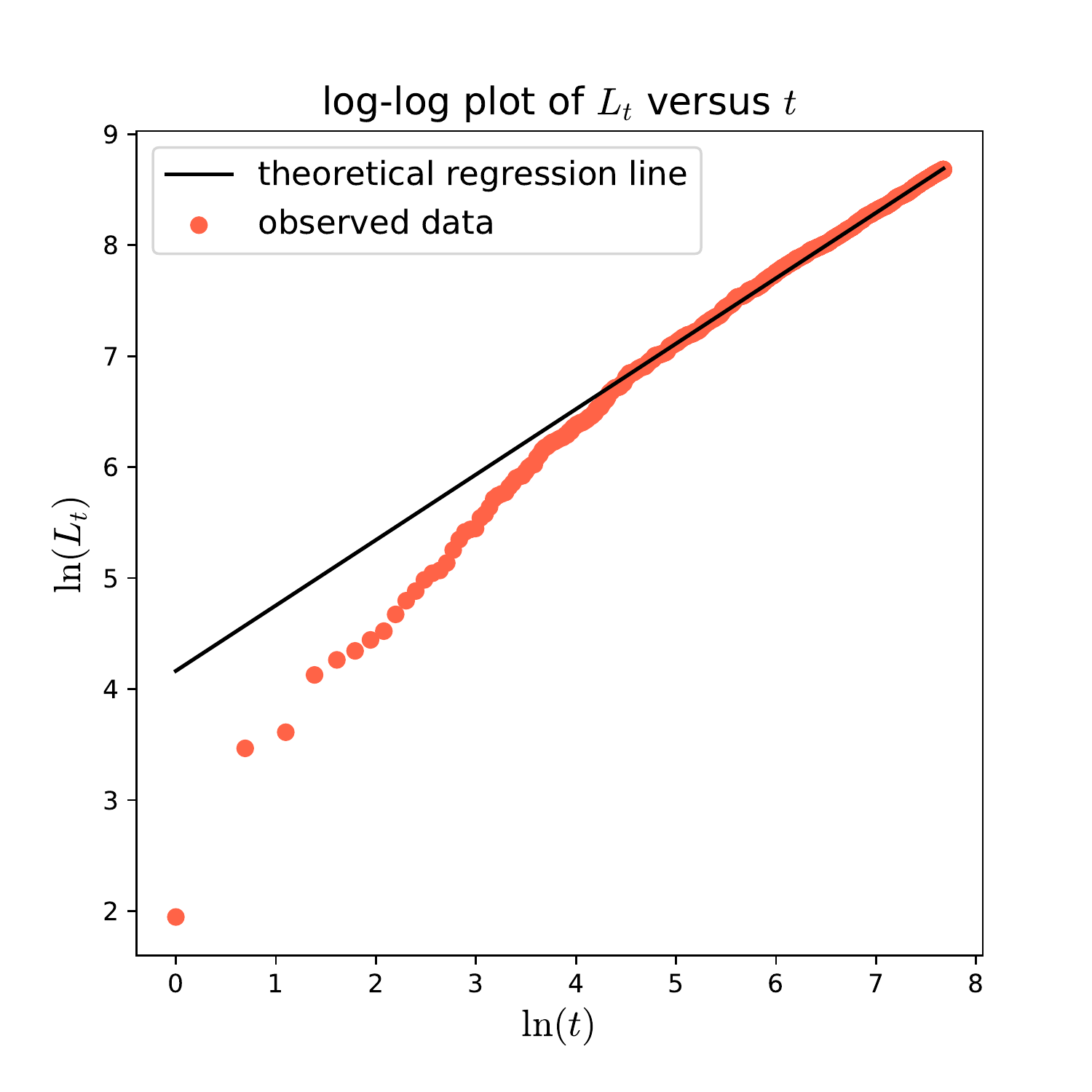}
\includegraphics[scale=0.4]{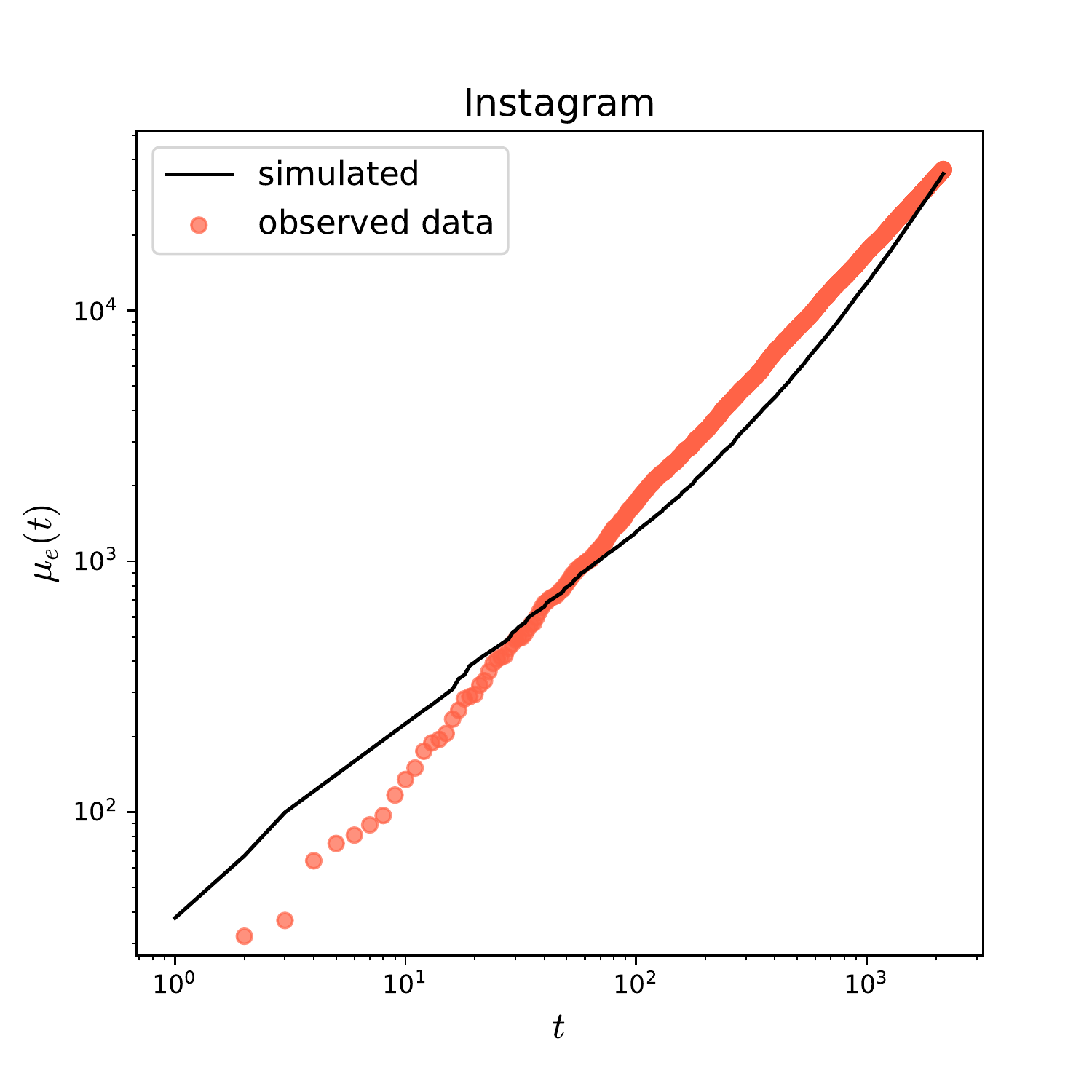}
\caption{\textbf{Instagram dataset.} Left: Plot of $\ln(L_t)$ as a
  function of $\ln(t)$, with the power-law trend. The red dots refer
  to the real data and the black line gives the theoretical regression
  line with slope $\widehat{\beta}$. Right: Asymptotic behavior of
  the number of edges in the actions-features network. Red dots refer
  to $e(t)$ of the real data, while the black line shows $\mu_e(t)$
  obtained by the model (averaging over $R = 100$ simulations).}
\label{fig:beta_plot_instagram}
\end{figure}

\indent For this application, we chose weights of type 5),
Subsection \ref{delta-weights}, 
that depend on an indicator related to the
underlying Instagram network. Precisely, we associate to each agent $i$
the variable $G_i$ defined as the number of agents $i$'s followers,
among those who were active during the crawling period and we set 
 
\begin{equation}\label{eq:weights_instagram}
W_{t,j,k} = W_{t} = e^{-G_{i(t)}},
\end{equation}
where $i(t)$ denotes the agent performing action $t$. Therefore the
inclusion probability for hashtag $k$ becomes
\begin{equation}
P_t(k) = \frac{\delta}{2} + 
(1-\delta)\ \frac{\sum_{j=1}^{t-1}F_{j,k}}{t}\; e^{-G_{i(t)}}\,,
\label{instagram_prob}
\end{equation}
where the average popularity of hashtag $k$ is exponentially
discounted by the factor $G_{i(t)}$. The decision to introduce such
kind of weights was driven by the following consideration.
A user with a very high number of followers identifies
  a person who is very popular on the social networks, an
  ``influencer'' in the extreme case. As a consequence, it may be
  reasonable to think that she/he is less affected by other people's
  posts and, consequently, less prone to use ``old'' hashtags.
For this user, the average popularity of $k$ in the inclusion
probability $P_t(k)$ should be less relevant.  On the contrary, a user
with a low number of followers may be more incline to follow the
current trends and the others' preferences and choices.
It is worthwhile to point out that in the
definition of the weights, we considered the number of followers of an
user as fixed to the value we observed at the end of the period of
observation (the crawling period).  In general, it may change in time,
depending on the changes in her/his network of ``virtual
friendships''. However, we assume it to be constant because of the
short time span considered.  \\

\begin{table}[tp]
\centering
\begin{tabular}{lcccc}
\toprule
$p$ & $\widehat{p}$ & $\overline{p}$ & $MSE(p)$\\
\midrule
$\alpha$ & 37.895 & 37.843 &  4.516 \\
$\beta$ & 0.5897 & 0.5900 & $7.89 \cdot 10^{-5}$\\
$\delta$ with weights \eqref{eq:weights_instagram} & 0.0063 & 0.0062 
& $2.70 \cdot 10^{-8}$\\
\bottomrule
\end{tabular}
\caption{\textbf{Instagram dataset}: Estimation of the model parameters. 
The average values $\overline{p}$ and the parameters' $MSE$ are computed 
over $R=100$ realizations of the model. See Subsection \ref{methodology} 
for further details.}
\label{table:estimation_instagram}
\end{table}
\begin{table}[t]
\centering
\begin{tabular}{lccccccccccc}
\toprule
Matrix ($T=2151$) & $L_T$ & $\sigma_{L_T}$ & $\overline{O}_T$ & $\sigma_{O_T}$ & $\overline{N}_T$ & $\sigma_{N_T}$ \\
\midrule
real & $5890$ & & $14.23$ & & $2.74$ & \\
Weights \eqref{eq:weights_instagram} & $5885$ & $42.39$ & $\mathbf{13.56}$ & $0.24$ & $2.74$ & $0.02$\\
Weights = 1 & $5899$ & $73.00$ & $80.03$ & $5.20$ & $2.74$ & $0.04$ \\
\bottomrule
\end{tabular}
\caption{\textbf{Instagram dataset.} Comparison between real and
  simulated actions-features matrices by means of the indicators
  \eqref{indicatori-confronto}.  For the simulations, all the
  considered quantities have been averaged over $R = 100$ realizations
  of the model. We also include an estimate of the variations around
  the averaged values, through the computation of the sample standard
  deviations. See Subsection \ref{methodology} for further details.}
\label{table:comparison_instagram}
\end{table}

\begin{table}[tp]
\centering
\begin{tabular}{lcc}
\toprule
Weights \eqref{eq:weights_instagram} & $\overline{m}_1$ & $\overline{m}_2$ \\
\midrule
$k^* = 5890$ (all observed features) & {\bf 0.99} & 0.037 \\
$k^* = 100$ & {\bf 0.97} &  \\
$k^* = 250$ & {\bf 0.98} & \\
$k^* = 500$ & {\bf 0.98} & \\
\bottomrule
\end{tabular}
\\
\begin{tabular}{lcc}
\toprule
Weights = 1 & $\overline{m}_1$ & $\overline{m}_2$ \\
\midrule
$k^* = 5890$ (all observed features) & 0.97 & {0.037} \\
$k^* = 100$ & 0.63 & \\
$k^* = 250$ & 0.77 & \\
$k^* = 500$ & 0.86 & \\
\bottomrule
\end{tabular}
\caption{\textbf{Instagram dataset.} Comparison between real and
  simulated actions-features matrices by means of the indicators
  \eqref{indicatori-corrispondenze}.  The first row of each table
  evaluates the indicators on the whole matrix ($k^*=5890$), while the
  other rows show the results computing the indicator $\overline{m}_1$
  only on the first $k^*$ ($=100,\,250,\,500$) features. See
  Subsection \ref{methodology} for further details.}
\label{table:prediction_instagram_2}
\end{table}

\begin{table}[tp]
\centering
\begin{tabular}{lccc}
\toprule
Weights \eqref{eq:weights_instagram} & $\overline{m}_1^*$ & $\overline{m}_2^*$ 
\\
\midrule
$T^* = 1613$ and $k^* = 5890$ & 0.99 & 0.006 \\
$T^* = 1076$ and $k^* = 5890$ & 0.99 & 0.031 \\
$T^* = 538 $ \, and $k^* = 5890$ & 0.99 & 0.099 \\
\midrule
$T^* = 1613$ and $k^* = 250$ & 0.98 & \\
$T^* = 1076$ and $k^* = 250$ & 0.98 & \\
$T^* = 538 $ \, and $k^* = 250$ & 0.97  & \\
\bottomrule
\end{tabular}
\caption{\textbf{Instagram dataset.} Predictions on the
  actions-features matrix. The indicators
  \eqref{indicatori-predizione} are computed for different levels of
  information used as ``training set'': more precisely, the different
  $T^*$ correspond to 75\%, 50\% and 25\% of the set of the actions,
  respectively. Moreover, the indicator $\overline{m}_1^*$ is computed
  on the whole matrix ($k^*=5890$) and also taking into account only
  the first $k^*=250$ features. See Subsection 
  \ref{methodology} for further details.}
\label{table:prediction_instagram}
\end{table}

\indent The performed analysis follows the outline explained in
Subsection \ref{methodology} (for the simulated matrices, all the
considered quantities have been averaged over $R = 100$ realizations
of the model). We first estimate the model's parameters, obtaining the
results in Table \ref{table:estimation_instagram}.  The weighted
preferential attachment term \eqref{wpa-factor} plays an important
role, but slightly lower than  in the previous cases,
since the inclusion probability is obtained with $\delta=0.6\%$.
Figure \ref{fig:beta_plot_instagram} provides in the left panel a
log-log plot of the cumulative count of new features (hashtags) as a
function of time (see the red dots), that clearly shows a power-law
behavior. This agrees with the theoretical property of the model
stated in the Appendix, Subsec.~\ref{total-num-features}, according to
which the power-law exponent has to be equal to the parameter $\beta$.
This fact is checked in the plot by the black line, whose slope is the
estimated value of the parameter $\beta$.  The goodness of fit of our
model to the dataset has been evaluated through the computation of the
quantities \eqref{indicatori-confronto} and
\eqref{indicatori-corrispondenze}.  These results are shown in Table
\ref{table:comparison_instagram} and Table
\ref{table:prediction_instagram_2}. Table
  \ref{table:comparison_instagram} shows that our model is
perfectly able to reproduce the total number $L_T$ of features
observed at the end of the observation period $T$, as well as the
average number of new features $\overline{N}_T$ in both the 
considered cases.   
The average number of old features (i.e.~the quantity
$\overline{O}_T$) shows a good agreement with the observed
quantity in the case of the model with the chosen weights, 
contrarily to the model with all the weights 
equal to one for which we obtain a much higher value. 
Table \ref{table:prediction_instagram_2} also indicates that 
the model with the chosen weights shows a better
performance than the one with all the weights equal to one. More
precisely, the values obtained for the indicator $\overline{m}_2$ are
almost the same for both cases (the average error on the total number
of arrived features is around $4\%$); while the most significant
differences are in the values of the indicator
$\overline{m}_1$. Indeed, for the model with the chosen weights, the
computed values of $\overline{m}_1$ ranges from $97\%$ to $99\%$,
pointing out that a high percentage of the entries in the
actions-features matrix have been correctly inferred by the model. The
differences are more evident when we select the first $k^*$ features:
indeed, with the chosen weights we succeed to infer the values of at
least $97\%$ of the entries; while with all the weights equal to $1$
the percentage remains under $86\%$. This means that the major
difference in the performance of the different considered weights is
in the first features, that are those for which the preferential
attachment term is more relevant.  In Table
\ref{table:prediction_instagram} we evaluate the predictive power of
the model with the chosen weights: we estimate the parameters of the
model only on a subset of the observed actions, respectively the
$75\%$, $50\%$ and $25\%$ of the total observations; we then predict
the features for the ``future'' actions $\{T^*+1, \dots, T\}$ and
compare the predicted and observed results by means of the indicators
in \eqref{indicatori-predizione} over the whole set of features and
only on a portion of it. The indicator $\overline{m}_1^*$ ranges from
$97\%$ to $99\%$.  Finally, in the right panel of Figure
\ref{fig:beta_plot_instagram}, we provide the asymptotic behavior of
the number of edges in the actions-features network: more precisely,
the red dots represent the total number $e(t)$ of edges observed in
the real actions-features matrix at each time-step; while the
continuous black line shows the mean number $\mu_e(t)$ of edges
obtained averaging over $R = 100$ simulations of the model with the
chosen weights.

\subsection{Summary of the results}
\label{discussion}
We here summarize the major findings of the three considered
applications.\\ 

\indent In all the three cases we selected the weights depending on
a fitness variable. In the first two applications (IEEE and arXiv),
the fitness variable measures the ``ability'' of the agents (authors)
to transmit the features (keywords) of their actions
(publications). In the third application (Instragram) the fitness
variable quantifies the ``inclination'' of the agents (users) to
follow the features (hashtags) of the previous actions (posts). From
the performed analyses of the actions-features bipartite networks, we
get the following main common issues for the three applications:
\begin{itemize}
\item The preferential attachment rule plays a relevant role in the
  formation of the actions-features network, because of the small
  estimated values obtained for the parameter $\delta$. In particular, 
  in the first two applications, the estimated value of $\delta$ is 
  very close to zero.
\item The considered indicators \eqref{indicatori-confronto},
  \eqref{indicatori-corrispondenze} and \eqref{indicatori-predizione}, 
  and the plots regarding the behaviors along time of the total number
  of observed features $L_t$ and the total number of edges $e(t)$ show
  a good fit between the model with the chosen weights and the real
  datasets. In particular, the power-law behavior of $L_t$ perfectly
  matches the theoretical one with the estimated parameter $\beta$ as
  the power-law exponent, and a high percentage of the entries of the
  actions-features matrix is successfully inferred with the
  model. Moreover, a good performance is also obtained when making
  a prediction analysis, i.e.~testing the percentage of the entries
  that are successfully recovered by the model providing it with
  different levels of information.
\item With respect to the ``flat weights'', i.e. all weights equal to
  $1$, the chosen weights guarantee a better agreement with the real
  actions-features matrices. Among the indicators in
    \eqref{indicatori-confronto}, the one that mostly put in evidence
    this fact is $\overline{O}_T$. Moreover, the difference in the
    performance of the model with different weights is also
    particularly evident when we consider a subset of the overall set
    of observed features for the computation of the indicators
    $\overline{m}_1$ in \eqref{indicatori-corrispondenze} and
    $\overline{m}_1^*$ in \eqref{indicatori-predizione}.  Indeed, the
    first features are those for which the preferential attachment
    term is more relevant.
\end{itemize}

\section{Discussion and conclusions}
\label{conclusion}

In this work we have presented our contribution to the stream of
literature regarding stochastic models for bipartite networks
formation. With respect to the previous publications, our paper
introduces some novelties. First of all, given a system of agents, we
are not interested in modeling the process of link formation between
the agents themselves, we instead define a model that describes the
activity of the agents, studying the behavior in time of agents'
actions and the features shown by these actions. This issue allows to
amplify the range of possible applications, since we only assume to
know the chronological order in which we observe the agents' actions,
and not the order in which the agents arrive. Second, we extend the
concept of ``preferential attachment with weights''
\cite{Bianconi2001,Bianconi2001a} to this framework. The weights can
have different forms and meanings according to the specific setting
considered and play an important role since the probability that a
future action shows a certain feature depends, not only on its
``popularity'' (i.e.~the number of previous actions showing the
feature) as stated by the preferential attachment rule, but also on
some characteristics of the agents and/or the features themselves. For
instance, the weights may give information regarding the ``ability''
of an agent to transmit the features of her/his actions to the future
actions, or the ``inclination'' of an agent to adopt the features
shown in the past. \\ \indent Summarizing, we first provide a full
description of the model dynamics and interpretation of the included
parameters and variables, also showing some theoretical results
regarding the asymptotic properties of some important
quantities. Moreover, we illustrate the necessary tools in order to
estimate the parameters of the model and we consider three different
applications.  For each of them, we evaluate the goodness of fit of
the model to the data by checking the theoretical asymptotic
properties of the model in the real data, by comparing several
indicators computed both on the real and simulated matrices, as well
as testing the ability of the model as a predictive instrument in
order to forecast which features will be shown by future actions.  All
our analyses point out a very good fit of our model and a very good
performance of the adopted tools in all the three considered
cases.\\ \indent Our model and the related analysis have been able to
detect some interesting aspects that characterize the different
examined contexts. In the first two applications (IEEE
  and arXiv) we examined the publications in the scientific areas of
  Automatic Driving and of High Energy Physics (briefly Hep-Th) and we
  took into account two kinds of fitness variables for the authors:
  one based on the number of publications and the other based on the
  number of collaborators. Our study reveals that, for Hep-Th, both
  the number of publications of an author and the number of her/his
  collaborators are able to provide a good agreement with real data,
  while, for Automatic Driving, we found a better performance of the
  model with the weights based on the number of publications. Probably
  this difference is due to the fact that, while the Physics of High
  Energies is quite an old subject in which different branches
  developed, Automatic Driving is a much younger, and so limited,
  research area. (Indeed, the observed values of $T$ and $L_T$, that
  is the number of publications and the number of keywords in the
  considered period, for the Automatic Driving are much smaller than
  the ones observed for Hep-Th. The indicator $\overline{N}_T$ also 
  suggests that Automatic Driving is a much younger research field 
  than Hep-Th, since the observed value for the former is much greater 
  than the one for the latter.)
The behavior of on-line social networks is completely different: we
examine the dataset of Instagram, with posts considered as actions and
hashtag as features. Indeed, we saw that the less followers a user has
the higher the number of ``old'' hashtag used. This could be related
to the fact that less popular users tend to re-use many ``old''
hashtags in order to increase their visibility, while highly famous
users do not feel the need of improving their popularity in this way
and focus on few ``old'' hashtags. Indeed, this behavior show a
completely different role of the ``on-line followership'' relations
respect to coauthorships: while collaborations incentive the usage of
a high number of existing features, the number of followers takes to a
limited usage of existing hashtags.

\section*{Acknowledgements}
This work of FS was supported by the EU projects CoeGSS
(Grant No. 676547), Openmaker (Grant No. 687941), SoBigData (Grant
No. 654024). CB and IC are members of the
Italian Group ``Gruppo Nazionale per l'Analisi Matematica, la
Probabilit\`a e le loro Applicazioni'' (GNAMPA) of the Italian
Institute ``Istituto Nazionale di Alta Matematica'' (INdAM). 
IC and FS acknowledge support from the Italian 
``Programma di Attivit\`a Integrata'' (PAI), project ``TOol for 
Fighting FakEs'' (TOFFE) funded by IMT School for Advanced 
Studies Lucca.

\section*{Competing interests statement}
The authors have no competing interests to declare.

\bibliographystyle{unsrt}
\bibliography{main_body_more_recent}

\newpage
\appendix
\section{Appendix}
\label{appendix}

In the appendix we collect all the technical results and details that,
for the sake of simplicity, have not been included in the main body of
the work.  Specifically, in Subsection \ref{appendix-asymptotics} we
describe the asymptotic behavior of the total number of features along
time and we show some analytical findings regarding the asymptotic
behavior of the mean number of edges in the actions-features bipartite
network; in Subsection \ref{estimation} we provide some statistical
tools in order to estimate the parameters of the model; finally, in
the last Subsection \ref{methodology}, we illustrate the indicators
used in order to analyze the three considered real datasets (arXiv,
IEEE, Instagram).

\subsection{Some asymptotic results for the model}
\label{appendix-asymptotics}
We here illustrate some asymptotic properties of the model. 

\subsubsection{Asymptotic behavior of the total number of features}
\label{total-num-features}
The random variable $L_t = \sum_{j = 1}^t N_j$, that represents the
total number of features present in the system at time-step $t$, has
the following asymptotic behaviors as $t\to +\infty$:
\begin{itemize}
\item[a)] for $\beta = 0$, we have a logarithmic behavior of $L_t$,
that is $L_t /\ln(t)\to \alpha$ almost surely; 
\item[b)] for $\beta \in (0, 1]$, we obtain a power-law behavior, 
i.e.~$L_t/t^{\beta}\to \alpha/\beta$ almost surely. 
\end{itemize}
The proof of these two statements is exactly the same as in
\cite{CRIMALDI2017}, since the weights do not affect $L_t$.

\subsubsection{Asymptotic behavior of the mean number of edges in the 
actions-features network}
\label{mean_num_edges}
\begin{figure}[t]
\centering
\includegraphics[scale=0.36]{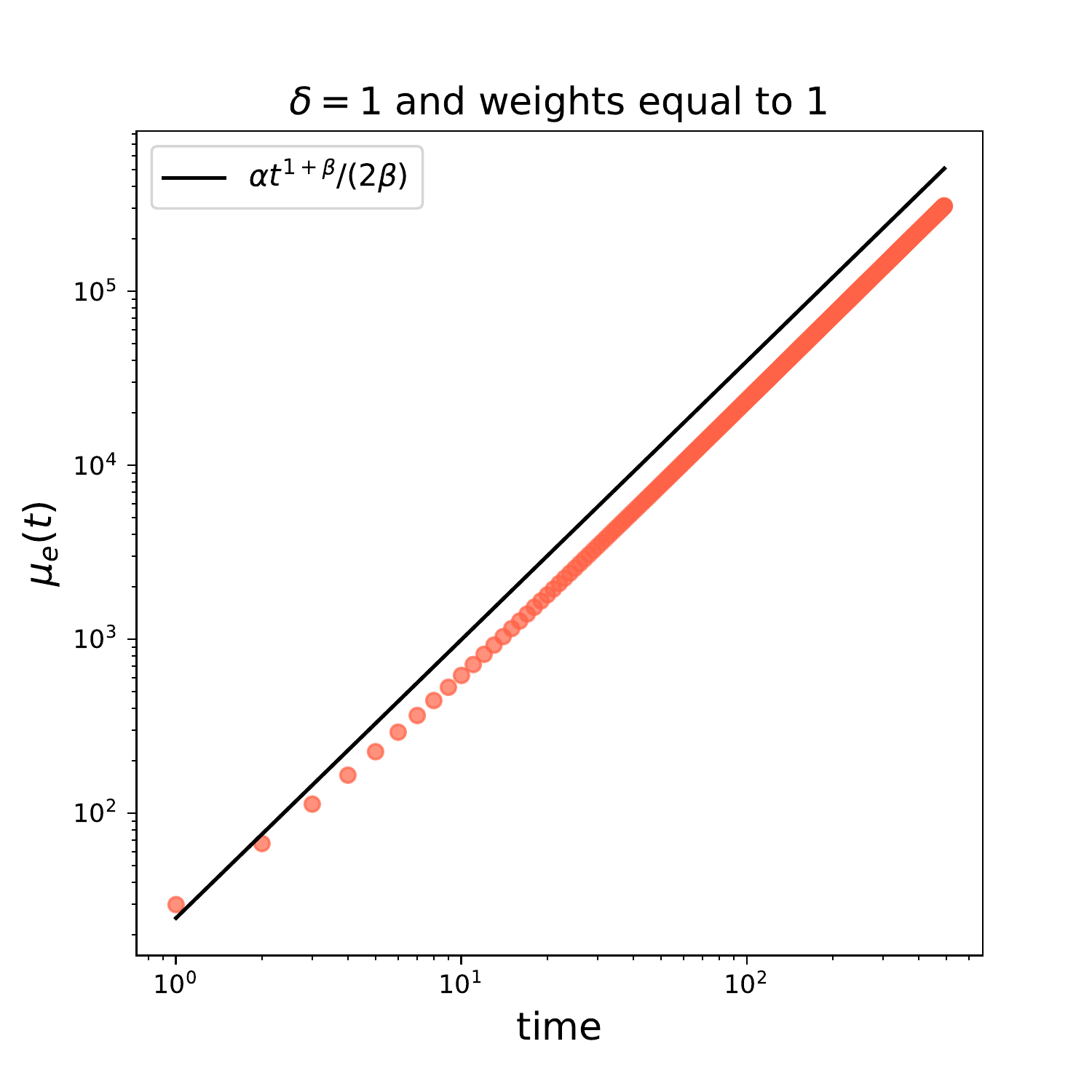}
\includegraphics[scale=0.36]{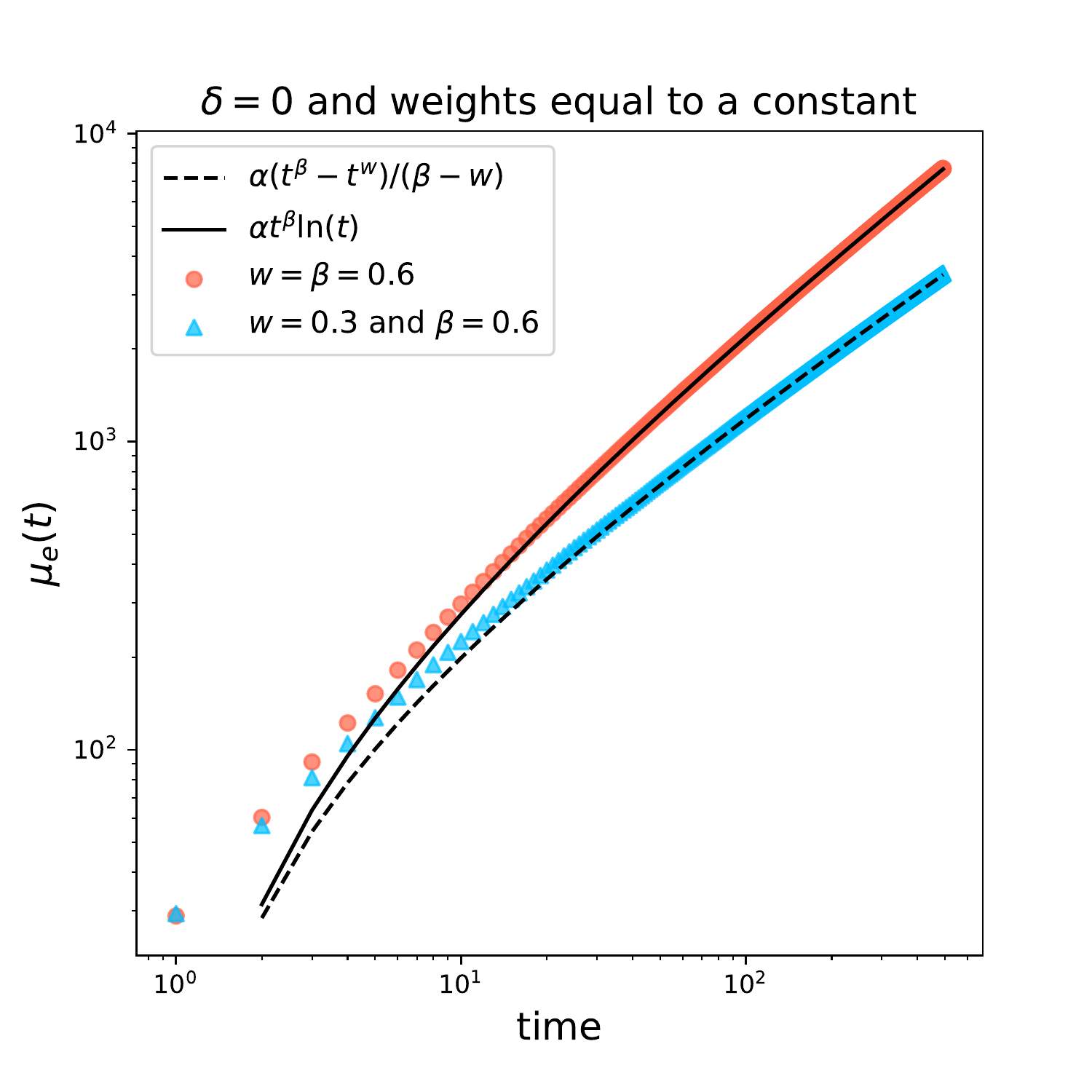}
\includegraphics[scale=0.36]{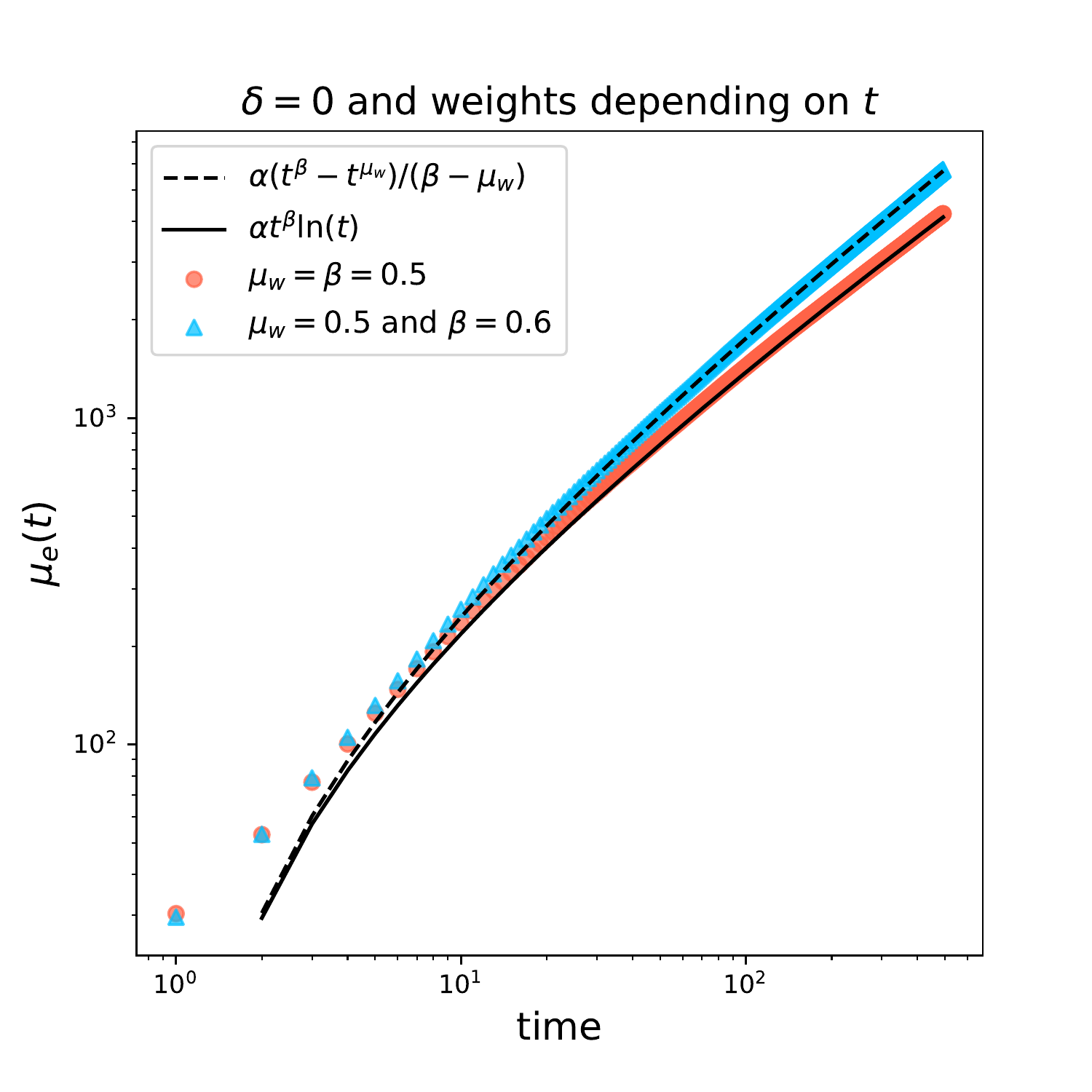}\\
\includegraphics[scale=0.36]{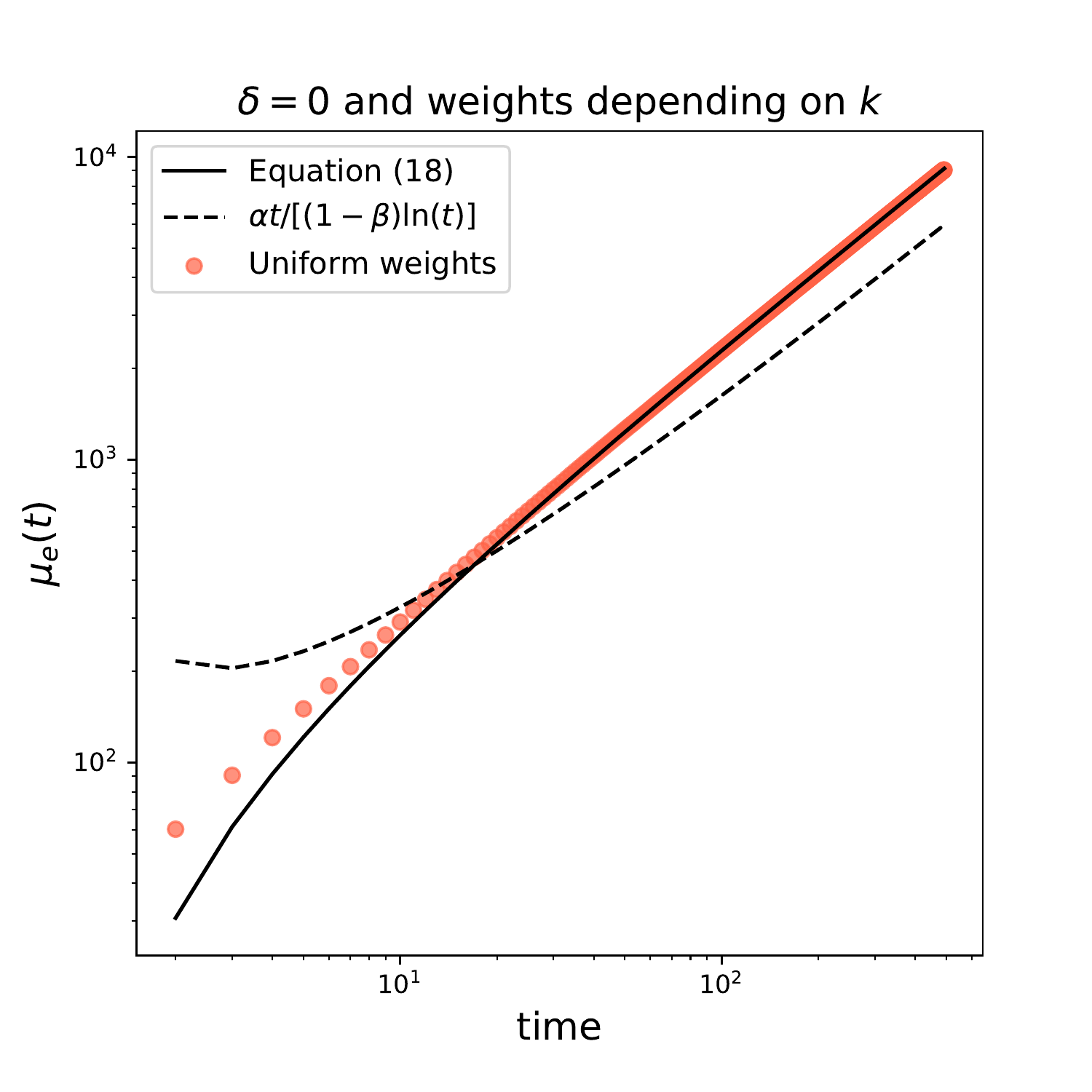}
\includegraphics[scale=0.36]{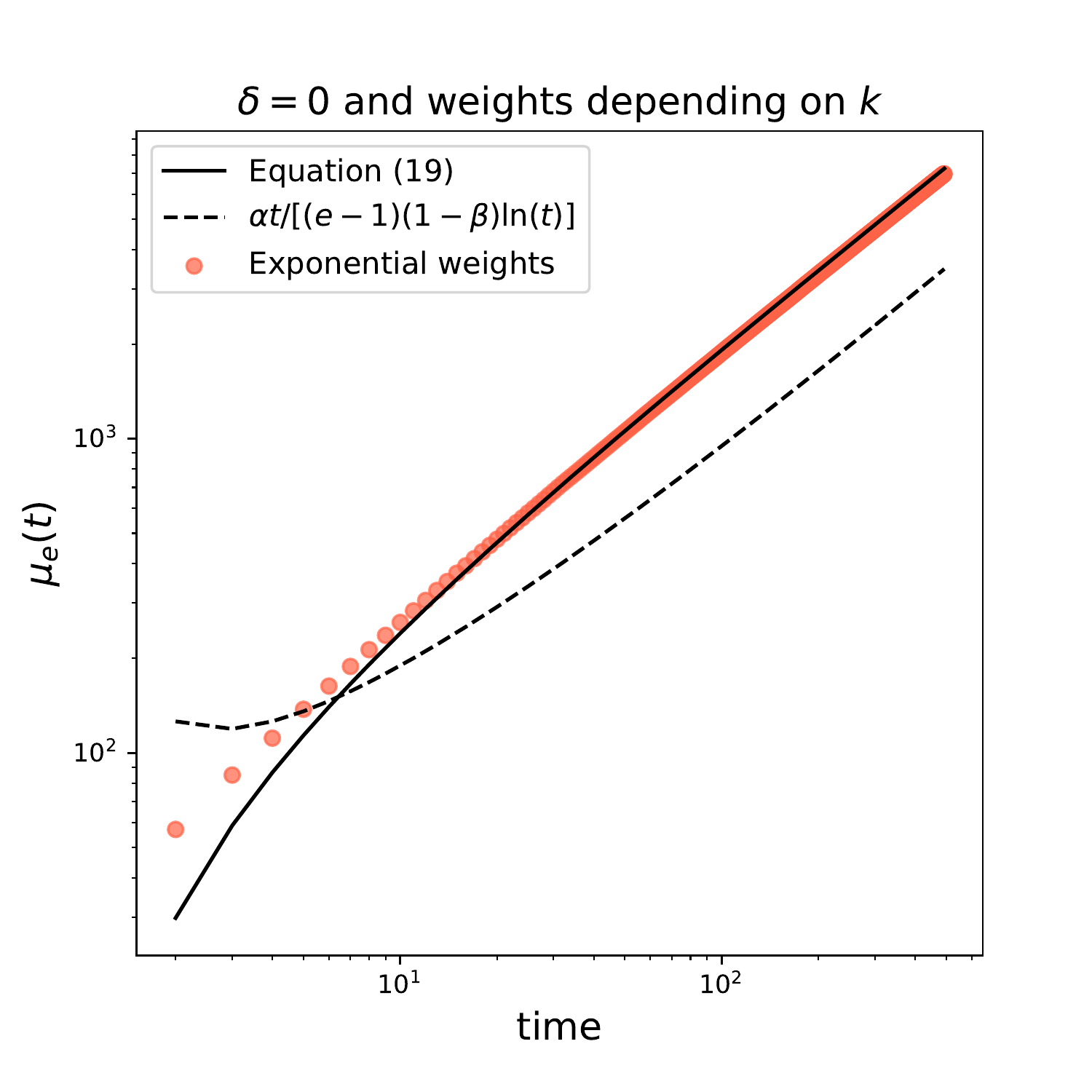}
\caption{Evolution of $\mu_e(t)$, i.e.~the mean number of edges along
  time. From top-left to bottom-right we have the cases: $\delta = 1$
  and weights equal to $1$; $\delta = 0$ and weights equal to a
  constant $w$ (the blue triangles represent the case with $w\neq
  \beta$, while the red dots show the case $w=\beta$); $\delta = 0$
  and weights depending only on $t$ with uniform distribution on
  $[0,1]$ (the blue triangles show the case with the mean value $\mu_W
  \neq \beta$, while the red dots describe the case with $\mu_W =
  \beta$); $\delta = 0$ and weights depending only on $k$, considering
  the two different distributions of the provided examples (uniform
  and truncated exponential distribution) for the weights: the
  continuous lines refer to the values of the integrals
  \eqref{eq:uniform_eta_magik} and \eqref{eq:exponential_eta},
  respectively, while the dashed lines show the final
  approximations. All simulations have been performed with $\alpha =
  30$ and $\beta=0.6$ (unless otherwise specified in the legend).}
\label{fig:mu_e}
\end{figure}

We here analyze the asymptotic behavior, as $t\to +\infty$, of
$\mu_e(t)=E[e(t)]$, where $e(t)$ is the total number of edges in the
actions-features network at time-step $t$, that is the total number of
ones in the matrix $F$ until time-step $t$. A first remark is that we
have
\begin{equation}\label{eq-e}
e(t)=\sum_{u=1}^t\sum_{k:\, T_k=u }d_k(t),
\end{equation}
 where we denote by
$T_k$ the arrival time-step of feature $k$ and 
\begin{equation}\label{eq-d_k}
d_k(t)=\sum_{j=1}^t F_{j,k}=1+\sum_{j=T_k+1}^t F_{j,k}
\end{equation}
 is the degree of feature $k$ at time-step $t$.  Hence, we can write
\begin{equation}\label{eq-base}
E[e(t)| T_k\;\forall k\,\mbox{with } T_k\leq t ]=
\sum_{u=1}^t\mbox{card}(k:\, T_k=u)E[d_{k}(t)|T_k=u]=
\sum_{u=1}^t N_u E[d_{k}(t)|T_k=u],
\end{equation}
where we recall that $N_u$ is Poi$(\lambda_u)$-distributed with
$\lambda_u=\alpha/u^{1-\beta}$.  In the following subsections, we go
further with the computations in the two ``extreme'' cases $\delta=1$
and $\delta=0$ since the behavior for a general $\delta$ is a mixture
of the two behaviors in the extreme cases.  A graphical representation
of the evolution of $\mu_e(t)$ in the considered cases is provided in
Figure \ref{fig:mu_e} (the values are averaged over a sample of
$R=100$ simulations).

\subsubsection*{The case $\delta=1$}
\label{delta_1}
In this case the inclusion probability of a feature $k$ at time-step
$t$ simply is $P_t(k)=\dfrac{1}{2}$.  Therefore, since \eqref{eq-d_k}, we have
$$
E[d_k(t)|T_k=t_k]=1+\dfrac{t-t_k}{2}\sim t/2.
$$ Hence, by \eqref{eq-base} and the above approximation, we can
approximate $\mu_e(t)$ by the quantity 
\begin{equation} 
\dfrac{t}{2}\sum_{u=1}^t\lambda_u=
\dfrac{\alpha t}{2}\sum_{u=1}^{t}u^{\beta-1}\sim 
\dfrac{\alpha t^{1+\beta}}{2\beta}.
\end{equation}

\subsubsection*{The case with $\delta=0$ and the weights equal to a constant}
Let us assume $\delta=0$ and $W_{t,j,k}$ equal to a constant $w\in
]0,1]$ for all $t,j,k$, so that the inclusion probability of a feature
    $k$ at time-step $t$  is
\begin{equation*}
P_t(k)=\dfrac{d_k(t-1)}{t}w.
\end{equation*}
Let us set  $\langle d_k(t)\rangle=E[d_k(t)|T_k=t_k]$ and observe that we have 
\begin{equation*}
\begin{split}
\langle d_k(t)\rangle=&1+w
\sum_{\tau=t_k+1}^t\dfrac{\langle d_k(\tau-1)\rangle}{\tau}\\
=&
1+w\Bigg[\sum_{\tau=t_k+1}^{t-1}
\dfrac{\langle d_k(\tau-1)\rangle}{\tau}+
\dfrac{\langle d_k(t-1)\rangle}{t}\Bigg]\\
=&1+w\sum_{\tau=t_k+1}^{t-1}\dfrac{\langle d_k(\tau-1)\rangle}{\tau}+
\dfrac{w}{t}\Bigg[1+w\sum_{\tau=t_k+1}^{t-1}
\dfrac{\langle d_k(\tau-1)\rangle}{\tau}\Bigg]\\
=&\Big(1+\dfrac{w}{t}\Big)
\Bigg[1+w\sum_{\tau=t_k+1}^{t-1}\dfrac{\langle d_k(\tau-1)\rangle}{\tau}\Bigg]\\
=&\cdots\\
=&\Big(1+\dfrac{w}{t}\Big)\Big(1+\dfrac{w}{t-1}\Big)\cdots
\Big(1+\dfrac{w}{t_k+1}\Big)\\
=&\dfrac{t_k!}{t!}\cdot(t+w)\cdot(t-1+w)\cdots(t_k+1+w)\\
=&\dfrac{t_k!}{t!}\cdot\dfrac{(t+w)\cdot(t-1+w)\cdots(t_k+1+w)
\cdot(t_k+w)\cdots(w+1)\cdot w}{(t_k+w)\cdots(w+1)\cdot w}.
\end{split}
\end{equation*}
Using the properties of the $\Gamma$-function, we can write 
\begin{equation}\label{eq:kk_eta}
\begin{split}
\langle d_k(t)\rangle=&\dfrac{t_k!}{t!}
\dfrac{\Gamma(t+w+1)!}{\Gamma(t_k+w+1)!}=
\dfrac{\Gamma(t_k+1)}{\Gamma(t+1)}
\dfrac{\Gamma(t+w+1)!}{\Gamma(t_k+w+1)!}
\sim
\Big(\dfrac{t}{t_k}\Big)^{w}.
\end{split}
\end{equation}
Therefore, by \eqref{eq-base} and the above approximation, we can
approximate $\mu_e(t)$ by the quantity 
\begin{equation}
\sum_{u=1}^t \lambda_u \dfrac{t^w}{u^w}
=\alpha t^w\sum_{u=1}^t u^{\beta-w-1}
\sim
\begin{cases}
\alpha t^\beta \ln(t)\quad&\mbox{if } w=\beta,
\\ 
\dfrac{\alpha}{\beta-w}(t^{\beta}-t^w)
\sim \dfrac{\alpha t^{\max\{w,\beta\}}}{|w-\beta|}
\quad&\mbox{if } w\neq \beta. 
\end{cases}
\end{equation}

\noindent{\bf Remark:} It is worthwhile to note that in the case of
weights of the form $W_{t,j,k}=W_t$ for all $t,j,k$, where the random
variables $W_t$ take values in $[0,1]$, are identically distributed
with mean value equal to $\mu_W$, and each of them is independent of
all the past until time-step $t-1$, we get for $\mu_e(t)$ the same
asymptotic behavior as above, but with $w=\mu_W$.

\subsubsection*{The case with $\delta=0$ and the weights depending only on $k$}
\label{weights-W_k}

Let us assume $\delta=0$ and $W_{t,j,k}=W_k$ for all $t,j,k$, where
the random variables $W_k$ take values in $[0,1]$, are independent and
identically distributed with probability density function $\rho$, and
each of them independent of the arrival time-step $T_k$ of the feature.
Moreover, we focus on the case $\beta<1$, that is more interesting
then the case $\beta=1$. In this case the inclusion probability is
\begin{equation*}
P_t(k)=\dfrac{d_k(t-1)}{t}W_k.
\end{equation*}
Using the same computations done above, we get  
\begin{equation*}
E[d_k(t)|T_k=t_k,\, W_k]
\sim
\Big(\dfrac{t}{t_k}\Big)^{W_k}
\end{equation*}
and so we can approximate $E[d_k(t)|T_k=t_k]$ by $\int_0^1
\left(\dfrac{t}{t_k}\right)^{w}\rho(w)\,dw$. Hence, using
\eqref{eq-base}, we can approximate $\mu_e(t)$ by 
\begin{equation}\label{integral}
\begin{split}
&\sum_{u=1}^t\lambda_u\int_0^1\left(\dfrac{t}{u}\right)^w\rho(w)dw
=\int_0^1 t^w \sum_{u=1}^t \lambda_u u^{-w}\rho(w)dw=
\\
&\alpha\int_0^1 t^w \sum_{u=1}^t u^{-(w-\beta+1)}\rho(w)dw
=
\alpha t^{\beta}\int_0^1 \dfrac{t^{ w-\beta}-1}{ w-\beta}\rho(w)dw.
\end{split}
\end{equation}

Therefore the asymptotic behavior of $\mu_e(t)$ depends on the
asymptotic behavior of the above integral. In the sequel we analyze
the case of the uniform distribution and the one of the ``truncated''
exponential distribution. To this purpose, we employ the
Exponential integral
\begin{equation*}
\text{Ei}(y)=-\int_{-y}^{+\infty}\dfrac{e^{-x}}{x}dx=
\int_{-\infty}^y \dfrac{e^v}{v}dv,
\end{equation*}
which has the property $\lim_{y\rightarrow +\infty} \frac{e^{y}}{y \text{Ei}(y)}=1$.
\\

\noindent{\bf Example 1} ({\em Uniform distribution on $[0,1]$})\\ If
$\rho(w)=1,\,\forall w\in [0, 1]$ and equal to zero otherwise, we can
compute the above integral and approximate $\mu_e(t)$ by 
\begin{equation}\label{eq:uniform_eta_magik}
\alpha t^{\beta}\left\{\int_{-\beta\ln(t)}^{(1-\beta)\ln(t)} 
\dfrac{e^{v}}{v}dv
-\int_{-\beta}^{1-\beta}\dfrac{1}{v}dv\right\}
=
\alpha t^\beta 
\Big\{\text{Ei}[(1-\beta)\ln(t)]-\text{Ei}[-\beta\ln(t)]+
\ln\Big(\dfrac{\beta}{1-\beta}\Big)\Big\}
\end{equation}
Using the asymptotic properties of the Exponential integral, we find
that the above quantity behavies for $t\to +\infty$ as
$$\dfrac{\alpha t}{(1-\beta)\ln(t)}.$$

\noindent{\bf Example 2} ({\em Exponential distribution on
  $[0,1]$})\\ If $\rho(w)= e^{1-w}/(e-1)$ for $w\in [0,1]$ and equal
to zero otherwise, the computation of the above integral leads to the
approximation for $\mu_e(t)$ given by
\begin{equation}\label{eq:exponential_eta}
\begin{split}
&\dfrac{\alpha  e^{1-\beta}}{(e-1)}t^\beta
\left\{
-\int_{-\beta}^{1-\beta}\dfrac{e^{-x}}{x}dx
+\int_{-\beta(\ln(t)-1)}^{(1-\beta)(\ln(t)-1)}\dfrac{e^{v}}{v}dv
\right\}
\\
&=
\dfrac{\alpha  e^{1-\beta}}{(e-1)}t^\beta
\Big\{
\text{Ei}[\beta]-\text{Ei}[-(1-\beta)]+
\text{Ei}[(1-\beta)(\ln(t) -1)]-\text{Ei}[-\beta(\ln(t)-1)]
\Big\}.
\end{split}
\end{equation}
Using the asymptotic properties of the Exponential integral, we find
that the asymptotic behavior for $t\to +\infty$ of the above
quantity is given by
$$
\dfrac{\alpha t}{(e-1)(1-\beta)\ln(t)}.
$$

\clearpage

\subsection{Estimation of the model parameters}
\label{estimation}
We here provide some statistical tools in order to estimate the
parameters of the model introduced in Section \ref{model}.

\subsubsection*{The parameters $\alpha$ and $\beta$}
The parameters $\alpha$ and $\beta$ can be estimated using a maximum
likelihood method, that is maximizing the probability to observe 
$\{N_1 = n_1, N_2 = n_2, \dots, N_T = n_T\}$. 
Since all the random variables $N_t$ are assumed independent
Poisson distributed, we have
\begin{equation}\label{eq:like_alpha_beta}
P(N_1 = n_1, \dots, N_T = n_T) = P(N_1 = n_1)\prod_{t=2}^T P(N_t = n_t)= 
\hbox{Poi}(\alpha)\{n_1\}\prod_{t = 2}^T \hbox{Poi}\left(\lambda_t\right)\{n_t\}.
\end{equation}
Hence, we choose as estimates the pair $(\widehat{\alpha},
\widehat{\beta})$ that maximizes function \eqref{eq:like_alpha_beta},
or equivalently its log-likelihood expression
$$
\ln\left(\hbox{Poi}(\alpha)\{n_1\}\right)+ 
\sum_{t=2}^T\ln\left(\hbox{Poi}\left(\lambda_t\right)\{n_t\}\right).
$$
\noindent {\bf Remark:} From the result stated in Subsection
\ref{total-num-features} we get that $\ln(L_t)/ \ln(t)$ is a strongly
consistent estimator for $\beta$.  Indeed:
\begin{itemize}
\item[a)] if $\beta=0$, then we have $L_t\stackrel{a.s.}\sim \alpha\ln(t)$ as
  $t\to +\infty$, so $\ln(L_t)\stackrel{a.s.}\sim\ln(\alpha) +
  \ln(\ln(t))$ and hence $\ln(L_t)/\ln(t)\stackrel{a.s.}\to 0=\beta$;
\item[b)] if $\beta \in (0,1]$, then we have $L_t\stackrel{a.s.}\sim
  (\alpha/\beta)t^{\beta}$ as $t\to +\infty$, so
  $\ln(L_t)\stackrel{a.s.}\sim\ln(\alpha/\beta) + \beta \ln(t)$, and hence
  $\ln(L_t)/\ln(t)\stackrel{a.s.}\to \beta$.\\
\end{itemize}

\subsubsection*{The parameter $\delta$}
\indent An estimate for the parameter $\delta$ is obtained maximizing
the probability to observe the given biadjacency
matrix rows $\{F_1=f_1, F_2=f_2, \dots , F_T=f_T\}$. More precisely,
we have
\begin{equation}
\begin{split}
&P\left(F_1 = f_1, \dots, F_T = f_T\right)= 
P\left(F_1 = f_1\right)
\prod_{t=2}^T P\left(F_t = f_t | F_1, \dots, F_{t-1}\right)= \nonumber 
\\
&
P\left(N_1 = n_1\right) \prod_{t=2}^T 
P\left(F_{t,k} = f_{t,k} \; \hbox{for } k = 1, \dots, L_{t-1}, 
N_t = n_t | F_1, \dots, F_{t-1}\right)= \nonumber \\
&\hbox{Poi}\left(\alpha\right)\{n_1\}
\prod_{t = 2}^T \hbox{Poi}\left(\lambda_t\right)\{n_t\} 
\left\{\prod_{k = 1}^{L_{t-1}} P_t(k)^{f_{t,k}} \left(1-P_t(k)\right)^{1-f_{t,k}}\right\},
\end{split}
\end{equation}
where $P_t(k)$ and $\lambda_t$ are defined in \eqref{incl_prob} and
\eqref{lambda}, respectively.  Hence, we choose $\widehat{\delta}$
that maximizes $P(F_1 = f_1, \dots, F_T = f_T)$. Since many terms in
the previous equation do not depend on $\delta$, the problem
simplifies into the choice of the value of $\widehat{\delta}$ that
maximizes the following function
\begin{equation}\label{eq:like_delta}
\prod_{t = 2}^T \prod_{k = 1}^{L_{t-1}} P_t(k)^{f_{t,k}} \left(1-P_t(k)\right)^{1-f_{t,k}}
\end{equation}
or, equivalently, taking the logarithm, 
\begin{equation}
\label{likelihood}
\sum_{t = 2}^T \sum_{k = 1}^{L_{t-1}}f_{t,k}
\ln\left(P_t(k)\right) + (1-f_{t,k})\ln\left(1-P_t(k)\right).
\end{equation}
It is worthwhile to note that the expression of
the weights inside the inclusion probability \eqref{incl_prob}
may possibly contain a parameter $\eta$. In this case, we maximize the 
above functions with respect to $(\delta,\eta)$.

\subsection{General methodology for applications}
\label{methodology}

We here provide a detailed outline of the performed data cleaning 
procedures and analyses used for the three considered real datasets.

\subsubsection*{Data cleaning procedure}
For the arXiv and IEEE datasets, the data preparation procedure has
been carried out using the Python package
\textit{NodeBox}\footnote{\url{https://www.nodebox.net/code/index.php/Linguistics\#loading_the_library}},
that allows to perform different grammar analyses on the English
language. We use the library to categorise (as noun, adjective, adverb
or verb) each word in all title's or abstract's sentences, with the
final purpose of selecting nouns and adjectives only. Then, all
selected words are modified substituting capital letters with
lowercases and transforming all plurals into singulars, again using
the \textit{NodeBox} package. Finally, we also remove special words
such as ``study'', ``analysis'' or ``paper'', that may often appear in
the abstract text but are not relevant for the description of the
topic and for the purpose of our analysis. Authors names are similarly
treated. Indeed, from each name we replace capital letters with
lowercases and we modify it by considering only the initial letter for
each reported name and the entire surname. To make an example, names
such as ``Peter Kaste'' or ``P. Jacob'' are respectively transformed
into ``p.kaste'' and ''p.jacob''. One drawback of this kind of
analysis is that authors with more than one names who reported all of
them or just some in different publications cannot be
distinguished. Indeed, in this situation they would appear as
distinct. For example ``A. N. Leznov'', ``A. Leznov'' or ``Andrey
Leznov'' may probably identify the same person who reported
respectively two initials, one initial or the full name in different
papers. However, with this transformation they appear as two distinct
authors, since they are respectively represented by the abbreviations
``a.n.leznov'' and ``a.leznov''. Despite this fact, no
  further disambiguation is performed on the names, since it would be
  computationally very expensive.

\subsubsection*{Estimation of the model parameters}

\noindent We provide the estimated value of the parameters
$\alpha,\beta$ and $\delta$ of the model by means of the tools
illustrated in Section \ref{estimation}. For each parameter
$p\in\{\alpha,\beta,\delta\}$, we also give the averaged value
$\overline{p}$ of the estimates on a set of $R$ realizations (the
value $R$ is specified in each example) and the related mean squared
error $MSE(p)$. The detailed procedure works as follows: starting from
the estimated values $\widehat{\alpha}$, $\widehat{\beta}$ and
$\widehat{\delta}$ (and the observed chosen weights), we generate a
sample of $R$ simulated actions-features matrices and we estimate
again the parameters on each realization, obtaining the values
$\widehat{\alpha}_r$, $\widehat{\beta}_r$ and $\widehat{\delta}_r$,
for $r = 1, \dots, R$. We then compute, for each parameter
$p\in\{\alpha,\beta,\delta\}$, the average estimate $\overline{p}$
over all the simulations and the $MSE(p)$, as follows
\begin{equation}\label{valutazione-stime}
\overline{p} = 
\frac{1}{R}\sum_{r = 1}^{R} \widehat{p}_r\qquad
MSE(p) = 
\frac{1}{R}\sum_{r = 1}^{R}\left(\widehat{p}_r - \widehat{p}\right)^2 .
\end{equation}

\subsubsection*{Check of the asymptotic behaviors}

\noindent We consider the behavior of the total number $L_t$ of
observed features along the time-steps $t$ and we compare it with the
theoretical one of the model (see Subsection
\ref{total-num-features}). In particular, for each applications, we
verify that the power-law exponent matches the estimated parameter
$\beta$. Moreover, we consider the behavior of the total number
$e(t)$ of edges in the real actions-features network and we compare it
with the mean number $\mu_e(t)$ of edges obtained averaging over $R$
simulated actions-features networks.

\subsubsection*{Comparison between real and simulated matrices and relevance 
of the weights}  

\noindent We compare the real and simulated actions-features matrices 
on the basis of the following indicators:
\begin{equation}\label{indicatori-confronto}
\begin{split}
L_T &=
\hbox{total number of features exhibited by the observed $T$ actions},
\\
\overline{O}_T &= \frac{1}{(T-1)}\sum_{t = 2}^T O_t
\quad\hbox{with }
O_t=\sum_{k = 1}^{L_{t-1}} F_{t,k}
\\
\overline{N}_T &= \frac{1}{T}\sum_{t=1}^T N_t.
\end{split}
\end{equation}
For each action $t$, with $2\leq t\leq T$, the quantity $O_t$ is the
number of ``old'' features shown by action $t$ and $N_t=L_{t}-L_{t-1}$
is the number of ``new'' features brought by action $t$. The
indicators $\overline{O}_T$ and $\overline{N}_T$ provide the averaged
values overall the set of observed actions. These indicators are
computed for the real matrix, for the simulated matrix by the model
described in Section \ref{model} with the chosen weights and, in order
to evaluate the relevance of the weights inside the dynamics, we also
compute them considering all the weights equal to $1$.  In particular,
for the simulated matrices, the provided values are an average on $R$
realizations. Essentially, for each indicator 
$I \in \{L_T, \overline{O}_T, \overline{N}_T\}$ the tables provide 
the average quantity
$$\overline{I} = \frac{1}{R} \sum_{r = 1}^R I_r,$$ where the term
$I_r$ denotes the quantity $I$ computed on the $r$-th simulation of
the model. Moreover, we approximate the variations around the average
values $\overline{I}$ by computing the sample standard deviation on
$R$ realizations, as follows
$$\sigma_I =
\sqrt{ \frac{1}{R-1} \sum_{r = 1}^R \left(I_r - \overline{I}\right)^2}.
$$
Furthermore, in order to take into account also the
not-exhibited ``old'' features (i.e.~the zeros in the matrix $F$), we
check also the number of ``correspondences'', that is we compute the
following indicators:
\begin{equation}\label{indicatori-corrispondenze}
\overline{m}_1=\frac{1}{R}\sum_{r=1}^{R} m_1^{sim_r}
\quad\hbox{and}\quad
\overline{m}_2=\frac{1}{R}\sum_{r=1}^{R} m_2^{sim_r},
\end{equation}
where
\begin{equation*}
\begin{split}
m_1^{sim_r} &=\frac{1}{T-1} \sum_{t=2}^T  m_1^{sim_r}(t)
\;\hbox{with  }
m_1^{sim_r}(t)=
\frac{1}{\min(L_{t-1}^{re}, L_{t-1}^{sim_r},k^*)} 
\sum_{k=1}^{ \min(L_{t-1}^{re}, L_{t-1}^{sim_r},k^*)}  
\mathbb{I}_{\{F_{t,k}^{re}=F_{t,k}^{sim_r}\}}
\\
\mbox{and}
\\
 m_2^{sim_r} &= \frac{1}{T-1} \sum_{t=2}^T m_2^{sim_r}(t)
\;\hbox{with  }
m_2^{sim_r}(t)= 
\frac{|L_{t}^{re} - L_{t}^{sim_r}|}{L_{t}^{re}}.
\end{split}
\end{equation*}
In the above formulas, we use the apex abbreviation $re$ or $sim_r$ to
indicate whether the considered quantity is related to the real
matrix or the $r$-th realization of the simulated matrix,
respectively. The meaning of the above indicators is the following.
Given a realization $r$ of the simulated matrix, 
for a certain action
$t$, the quantity $m_1^{sim_r}(t)$ calculates the total number of
correctly attributed ``old'' features among the features in
$\{1,\dots, k^*\}$; while $m_2^{sim_r}(t)$ computes the relative error
in the total number of observed features. Then, $m_1^{sim_r}$ and
$m_2^{sim_r}$ are the corresponding averaged values overall the set of
observed actions, and $m_1$ and $m_2$ are the averaged values over the
$R$ realizations of the simulated matrix.  Values of $m_1$ and
$m_2$ respectively close to $1$ and $0$ indicate that a very high
fraction of features has been correctly allocated by our model and that
the relative error in the total number of observed features is very
low.

\subsubsection*{Predictive power of the model}  

\noindent We perform a prediction analysis on the actions-features 
matrix.  More precisely, once a time-step $T^*< T$ is fixed, we
estimate the model parameters on the ``training set'' corresponding to the
set of actions observed at $t = 1, \dots, T^*$.  We then employ
those estimates to simulate the dynamics of the actions-features
network related to the remaining set of actions at times $t = T^*+1,
\dots, T$. Finally, taking the features really observed for these last
actions as ``test set'', we evaluate the goodness of our predictions 
by computing the following indicators:
\begin{equation}\label{indicatori-predizione}
\overline{m}_1^*=\frac{1}{R}\sum_{r=1}^{R} m_1^{*,sim_r}
\quad\hbox{and}\quad
\overline{m}_2^*=\frac{1}{R}\sum_{r=1}^{R} m_2^{*,sim_r},
\end{equation}
where
\begin{equation*}
\begin{split}
m_1^{*,sim_r} &= 
\frac{1}{T-T^*} \sum_{t=T^*+1}^T  m_1^{*,sim_r}(t)
\;\hbox{with  }
m_1^{*,sim_r}(t)=
\frac{1}{\min(L_{t-1}^{re}, L_{t-1}^{sim_r}, k^*)} 
\sum_{k=1}^{ \min(L_{t-1}^{re}, L_{t-1}^{sim_r}, k^*)}  
\mathbb{I}_{\{F_{t,k}^{re}=F_{t,k}^{sim_r}\}}\\
\mbox{and}
\\
m_2^{*,sim_r} &= \frac{1}{T-T^*} \sum_{t=T^*+1}^T m_2^{*,sim_r}(t)
\;\hbox{with  }
m_2^{*,sim_r}(t)= 
\frac{|L_{t}^{re} - L_{t}^{sim_r}|}{L_{t}^{re}}.
\end{split}
\end{equation*}
In the above formulas, as before, we use the apex abbreviation $re$ or
$sim_r$ to indicate whether the considered quantity is related to the
real matrix or the $r$-th realization of the simulated matrix,
respectively. The meaning of the above indicators is the following.
Given a realization $r$ of the simulated matrix, for a certain action
$t$, with $T^*+1\leq t\leq T$, the quantity $m_1^{*,sim_r}(t)$
calculates the total number of correctly attributed ``old'' features
among the features in $\{1, \dots , k^*\}$, 
while $m_2^{*,sim_r}(t)$ computes the relative error in the total
number of observed features. Then, $m_1^{*,sim_r}$ and $m_2^{*,sim_r}$
are the corresponding averaged values over the ``test set'' of
actions, and $m_1^*$ and $m_2^*$ are the averaged values over the $R$
realizations of the simulated matrix.  Values of $m_1^*$ and $m_2^*$
respectively close to $1$ and $0$ indicate that, starting from the
observation of the first $T^*$ actions (the ``training set''), a very
high fraction of features has been correctly predicted by our model
and that the relative error in the total number of observed features
is very low.  \\

\indent Regarding the prediction, it is worthwhile to note that if the
weights chosen in the model do not depend on $t$, then for the
predictions it is not important to know the agents performing the
actions $T^*+1,\dots, T$, but it is enough to have complete
information about the actions in the ``training set'' $1,\dots, T^*$.
Otherwise, if the weights depend on $t$, we need to assume also the
knowledge of all the agents performing the actions at time-steps
$T^*+1,\dots, T$, in order to take the right weights in the simulation
of the model at each time-step $t=T^*+1,\dots, T$ and predict the
corresponding features.

\end{document}